\documentclass[onecolumn]{aa} 
\usepackage{graphicx}
\usepackage[usenames,dvipsnames]{color}
\usepackage{txfonts}
\usepackage{natbib}
\usepackage{ifpdf}
\ifpdf
 \usepackage[pdftex,linktocpage=true,unicode=true,colorlinks=true,citecolor=red,urlcolor=blue,pdfstartview=FitH]{hyperref}
\else
 \usepackage[linktocpage=true,unicode=true,colorlinks=true,citecolor=red,urlcolor=blue,pdfstartview=FitH]{hyperref}
\fi
\newcommand{\beq}{\begin{equation}}
\newcommand{\eeq}{\end{equation}}
\newcommand{\bea}{\begin{eqnarray}}
\newcommand{\ena}{\end{eqnarray}}

\newcommand{\ie}{{\it i.e.}}
\begin{document}
\ifpdf
 \DeclareGraphicsExtensions{.png,.pdf,.jpg}	
\else
 \DeclareGraphicsExtensions{.eps,.ps}	
\fi
\title{The GeV-TeV Galactic gamma-ray diffuse emission}
\subtitle{I. Uncertainties in the predictions of the hadronic component}
\author{
T.~Delahaye
\inst{1}
\and
A.~Fiasson
\inst{2}
\and
M.~Pohl
\inst{3,4}
\and
P.~Salati
\inst{5}
}

\institute{
Instituto de F\'isica Te\'orica UAM/CSIC
Universidad Aut\'onoma de Madrid
Cantoblanco, 28049 Madrid, Spain\\
\email{timur.delahaye@uam.es}
%
%
\and
LAPP, Universit\'e de Savoie, CNRS, BP110, F-74941 Annecy-le-Vieux Cedex, France\\
\email{fiasson@lapp.in2p3.fr}
\and
Institut f\"ur Physik und Astronomie, Universit\"at Potsdam,
Karl-Liebknecht-Strasse 24/25, 14476 Potsdam, Germany\\
\email{pohlmadq@gmail.com}
\and
DESY, Platanenallee 6, 15738 Zeuthen, Germany
\and
LAPTH, Universit\'e de Savoie, CNRS, BP110, F-74941 Annecy-le-Vieux Cedex, France\\
\email{salati@lapp.in2p3.fr}
}

\date{Received \today ;}

%

\abstract
%
{The Galactic $\gamma$-ray diffuse emission is currently observed in the
GeV-TeV energy range with unprecedented accuracy by the Fermi satellite.
Understanding this component is crucial because it provides a background to
many different signals, such as extragalactic sources or annihilating
dark matter. It is timely to reinvestigate how it is calculated and
to assess the various uncertainties that are likely to affect the accuracy
of the predictions.}
%
{The Galactic $\gamma$-ray diffuse emission is mostly produced above
a few GeV by the interactions of cosmic ray primaries impinging on
the interstellar material. The theoretical error on that component
is derived by exploring various potential sources of uncertainty.
Particular attention is paid to cosmic ray propagation. Nuclear cross
sections, the proton and helium fluxes at the Earth's position, the Galactic
radial profile of supernova remnants, and the hydrogen distribution
can also severely affect the signal.}
%
{The propagation of cosmic ray species throughout the Galaxy is described
in the framework of a semi-analytic two-zone diffusion/convection model.
The $\gamma$-ray flux is reliably and quickly determined. This allows 
conversion of the constraints set by the boron-to-carbon data into a theoretical
uncertainty on the diffuse emission.
New deconvolutions of the HI and CO sky maps are also used to get the
hydrogen distribution within the Galaxy.}
%
{The thickness of the cosmic ray diffusive halo is found to have a
significant effect on the Galactic $\gamma$-ray diffuse emission, while
the interplay between diffusion and convection has little influence on
the signal.
The uncertainties related to nuclear cross sections and to the
primary cosmic ray fluxes at the Earth are significant.
The radial distribution of supernova remnants along the Galactic
plane turns out to be a key ingredient.
As expected, the predictions are extremely sensitive to
the spatial distribution of hydrogen within the Milky Way.}
%
{Most of the sources of uncertainty are likely to be
reduced in the near future. The stress should be put (i) on better
determination of the thickness of the cosmic ray diffusive halo and
(ii) on refined observations of the radial profile of supernova remnants.}


\maketitle

%
\section{Introduction}
\label{sec:introduction}

%
The $\gamma$-ray sky at GeV-TeV energies is currently observed with unprecedented
accuracy by the Fermi satellite \citep[see for instance][]{Knodlseder:2010iz}.
Among the various fields of investigation, the Galactic diffuse emission
\citep{2009PhRvL.103y1101A} plays a very special role.
Modeling that component is actually crucial for extracting a residual isotropic
emission and may severely affect how the extragalactic $\gamma$-ray background
is derived from observations. Theoretical errors in the predictions
of the Galactic diffuse emission translate into systematic uncertainties
on the extragalactic background and indirectly affect the studies of blazars
and extragalactic active nuclei.
The Galactic diffuse emission is also a natural background to many different
signals, such as point sources or annihilating dark matter. Weakly interacting
massive particles are actually one of the favored candidates for astronomical
dark matter. They are expected to continuously annihilate within the Milky Way
and to produce high-energy photons swamped inside the conventional Galactic
diffuse emission. The latter needs to be accurately determined since it is the
astrophysical background from which a possible dark matter contribution may have
to be disentangled. The Galactic diffuse emission bears also upon
searches for axion like particles or decaying gravitinos.
It is thus timely to reinvestigate how it is calculated and to assess
the various uncertainties that are likely to affect the accuracy with
which it is derived.

\vskip 0.1cm
The hadronic component of the Galactic $\gamma$-ray diffuse emission
dominates the inverse Compton and the Bremsstrahlung mechanisms
above a few hundred MeV~\citep[see for instance][]{1977ApJ...212...60S,2010ApJ...722L..58S,2011arXiv1101.1381S} supported by EGRET~\citep{1997ApJ...481..205H} and Fermi~\citep{2009PhRvL.103y1101A} observations. It is produced by the interactions of high-energy
cosmic ray protons and helium nuclei impinging on interstellar gas
and is, as such, a probe of the distribution of hydrogen within the
Galactic plane.
It is also extremely sensitive to the distribution of cosmic ray
primaries along the line of sight. Determining the hydrogen
density from the Galactic diffuse emission requires thus knowing how
protons and helium nuclei are distributed within the Galaxy. Cosmic
ray transport is a key ingredient.
In this article, it is described in the framework of a semi-analytic
two-zone diffusion/convection model, which has been extensively discussed
by \citet{Maurin:2001sj,Maurin:2002ua}. This allows fast and reliable
calculation of the $\gamma$-ray flux according to a procedure explained
in section~\ref{sec:theory}. Determining the hadronic component of the
Galactic diffuse emission only requires a minute of CPU time on an
ordinary PC.
Notice that in \citet{2010ApJ...722L..58S}, although six different CR propagation models are considered and their effect on the $\gamma$-ray flux are investigated, no systematic analysis is presented.
\vskip 0.1cm
A scan of the parameter space pertinent to cosmic ray transport and
compatible with the boron-to-carbon data has been performed
by \citet{Maurin:2001sj}, who find that many different propagation
models are allowed by observations.
In section~\ref{sec:cr_propagation_error}, we take advantage of our
semi-analytic derivation of the $\gamma$-ray flux to quantify the
theoretical uncertainties arising from cosmic ray propagation. The
effect of each parameter is scrutinized.
The normalization and spectral index of the space diffusion coefficient
are varied without much repercussion on the signal. Convection comes into
play at low energies where it competes with diffusion. The overall
effect on the $\gamma$-ray flux is moderate, though.
More important are the dimensions of the cosmic ray diffusive halo whose
radial and vertical extensions are varied.
The larger the diffusive halo, the more abundant the hydrogen illuminated
by cosmic ray protons and helium nuclei, and the stronger the diffuse emission.
Variations in the $\gamma$-ray flux as large as 30\% are found towards some
directions when the half thickness $L$ of the diffusive halo is increased
from 1 to 4~kpc.
\vskip 0.1cm
The uncertainties related to the nuclear cross sections and to the proton
and helium fluxes measured at the Earth's position are investigated in
sections~\ref{sec:sigma} and \ref{sec:phi_proton_alpha_at_earth}. 
They are found to depend on the energy.
In the former case, they amount to 33\% at 1~GeV, reach a maximum
of 54\% at 4.5~GeV, and decrease down to 20 to 30 \% above 100~GeV.
In the latter case, they increase with energy to reach a value of $\pm$37\% at 1~TeV.
We also derive
the local effective $\gamma$-ray emissivity per hydrogen atom and compare
our results to the recent Fermi measurements \citep{Abdo:2009ka}.
In section~\ref{sec:f_SNR_radial}, we investigate how the distribution of
primary cosmic ray sources along the Galactic plane affects the signal.
We find that the radial profile of supernova remnants has a significant
effect on the $\gamma$-ray diffuse emission, with variations as large as
50\% towards the Galactic center and 70\% in the opposite direction.
%
%
We finally gauge the sensitivity of the
Galactic $\gamma$-ray diffuse emission to the spatial distribution of
hydrogen in section~\ref{sec:galactic_H}. The former is a probe of the latter and we find, as expected,
that predictions depend significantly on the HI and CO three-dimensional
(3D) maps selected for the calculation, as well as on the assumptions on
the $X_{\rm CO}$ conversion factor.
We base our predictions on new deconvolutions of the HI and CO sky maps
\citep{2008ApJ...677..283P}, which we compare to the hydrogen distributions
provided by the GALPROP package~\citep{1998ApJ...509..212S}.
In section~\ref{sec:conclusions}, we summarize the results of our analysis
and outline the ingredients whose knowledge needs to be improved in order
to make the Galactic diffuse emission an efficient probe of the distribution
of hydrogen within the Milky Way.

%
\section{Semi-analytic derivation of the gamma-ray flux}
\label{sec:theory}

To derive the hadronic component of the Galactic $\gamma$-ray diffuse
emission, we first express the flux at the Earth's position as a function of the
$\gamma$-ray emissivities of the nuclei of the interstellar medium (ISM).
The local effective emissivity ${\cal E}_{\rm eff}(\odot)$ of the ISM
per hydrogen atom is computed and compared to the value observed
recently by the Fermi collaboration \citep{Abdo:2009ka}. The discussion
of the uncertainties related to the $\gamma$-ray production cross
sections and to the primary cosmic ray (CR) fluxes is postponed to
sections~\ref{sec:sigma} and \ref{sec:phi_proton_alpha_at_earth}.
We then model the propagation of CR primaries within the
magnetic fields of the Milky Way. This allows us to compute the proton
and helium fluxes at any position $\mathbf{x}$ of the Galactic CR
diffusive halo once these fluxes have been measured at the Earth's position.
We finally explain how to speed up the calculation of the $\gamma$-ray flux
by a careful selection of the integrals to be performed.

%
\subsection{Gamma-ray emissivity of the interstellar medium}
\label{sec:emissivity}

%
Above a few GeV, most of the $\gamma$-ray diffuse emission is produced
by the interactions of CR protons and $\alpha$ particles impinging on
the nuclei of the ISM. This yields the differential $\gamma$-ray flux
at the Earth:
\beq
\Phi_{\gamma}(l,b,E) \, = \, {\displaystyle \int}_{\rm \!\!\! los} ds \;
\left\{
n_{\rm H} (\mathbf{x}) \, {\cal E}_{\rm H} (\mathbf{x},E) \; + \;
n_{\rm He}(\mathbf{x}) \, {\cal E}_{\rm He}(\mathbf{x},E) \; + \;
n_{\rm C} (\mathbf{x}) \, {\cal E}_{\rm C} (\mathbf{x},E) \; + \; \ldots
\; \equiv \;
{\displaystyle \sum_{\rm A}} \;
n_{\rm A} (\mathbf{x}) \, {\cal E}_{\rm A} (\mathbf{x},E)
\right\} ,
\label{flux_nA_EA}
\eeq
towards the direction defined by the Galactic longitude $l$ and latitude
$b$. The $\gamma$-ray energy is denoted by $E$.
The flux $\Phi_{\gamma}$ involves the convolution along the line
of sight (los) of the densities $n_{\rm A}$ of the various ISM
elements with their $\gamma$-ray emissivities ${\cal E}_{\rm A}$.
The position $\mathbf{x}$ is completely specified by $b$, $l$,
and the depth $s$ along the los.

\vskip 0.1cm
When irradiated by high-energy CR species, the nuclei of the ISM
shine in the $\gamma$-ray band. The rate at which a particular nucleus A
produces high-energy photons of energy $E$ per unit of energy and solid
angle defines the emissivity ${\cal E}_{\rm A}$. This rate is expressed
in units of photons s$^{-1}$ sr$^{-1}$ GeV$^{-1}$. In the case of hydrogen for
instance, the emissivity is given by the convolution of the impinging CR
proton and helium fluxes $\Phi_{p}$ and $\Phi_{\alpha}$ with the corresponding
differential $\gamma$-ray production cross sections and may be expressed as
\beq
{\cal E}_{\rm H}(\mathbf{x},E) \, = \,
{\displaystyle \int}_{\!\!\! T_{\! min}}^{+ \infty} dT \;
\left\{
{\displaystyle \frac{d{\sigma}}{dE}}
\left(
p[T] + {\rm H} \to \gamma[E]
\right) \times \Phi_{p}(\mathbf{x},T)
\; + \;
{\displaystyle \frac{d{\sigma}}{dE}}
\left(
\alpha[T] + {\rm H} \to \gamma[E]
\right) \times \Phi_{\alpha}(\mathbf{x},T)
\right\}.
\label{def:emissivity_H}
\eeq
The integral~(\ref{def:emissivity_H}) runs over the kinetic energy
\emph{per nucleon\/} $T$ of the projectiles.

\vskip 0.1cm
The differential cross section
${d{\sigma}}/{dE}(p[T] + H \to \gamma[E])$ accounts for the
interaction of a CR proton with kinetic energy $T$ colliding upon
a hydrogen nucleus of the ISM to yield a photon with energy $E$.
At low $T$, typically below 2 to 3 GeV, one of the protons is excited
into a $\Delta$ resonance, which subsequently decays back into a proton
and a neutral pion \citep{Stecker:1970sw}, the latter yielding a pair
of photons. The threshold for the production of the $\Delta$ resonance
translates into the condition
\beq
T \geq {\displaystyle \frac
{{\left( m_{\Delta} \, + \, 3 m_{p} \right)}{\left( m_{\Delta} \, - \, m_{p} \right)}}
{2 m_{p}}} \geq {T_{\! min}} \, = \, 0.28 \; {\rm GeV} ,
\eeq
where the minimal value of the $\Delta$ resonance mass has been
set equal to $m_{\Delta} = m_{p} + m_{\pi^{0}}$.
%
At higher energies, deep inelastic scattering comes into play, and
the isobaric approach needs to be gradually replaced with a scaling
model accounting for quark interactions, hadronization, and pion
production. 
The first parameterization of pion production in proton-proton collisions has been proposed by~\citet{1973ApJ...185..499S} who compiled collider data, collected from 1956 to 1972, with beam energy up to $\sim$~1~TeV.
The pion channel alone contributes 80\% of the total
photon yield.
Other mesons, such as the charged and neutral kaons, come into play
as do a few baryons, such as the $\Lambda$ and $\Sigma$ states, which
are also produced in hadronic showers.
Photons may also be directly generated in hadronic collisions.
This process contributes even more than the other kaon and baryon
channels over almost the entire energy domain and contributes 10
to 20\% to the total photon yield.
Based partially on the data collected by~\citet{1973ApJ...185..499S},
\citet{Kamae:2006bf} propose a parameterization of the
differential $\gamma$-ray production cross section in proton-proton
collisions which is quite reliable below a projectile energy
$E_{p} = m_{p} + T$ of order 50 GeV. Above that energy, the
inclusive cross section for pion production, which
\citet{Kamae:2006bf} have borrowed from \citet{Blattnig:2000zf},
is significantly overestimated.
Our article is based on the recent analysis by \citet{Huang:2006bp}.
The \citet{Kamae:2006bf} parameterization of the differential
$\gamma$-ray production cross section in proton-proton collisions
is used below a proton energy $E_{p}$ of 2.5 GeV. Above $E_{p} = 20$
GeV, the photon production is given by the DPMJET-III Monte Carlo
simulation of proton-proton, proton-nucleus and nucleus-nucleus
collisions \citep{Roesler:2000he,Roesler:2001mn}. The DPMJET-III
event generator is a very reliable code at high energies.
The requirement of unitarity, for instance, is built in, and no cross
section becomes negative regardless of the energy.
Although this code extrapolates low-energy data to higher energies, it has been
used at CERN to calculate the shielding of the detectors and colliders.
Between 2.5 and 20 GeV,
\citet{Huang:2006bp} interpolate linearly from the \citet{Kamae:2006bf}
parameterization and the DPMJET-III results.
Because \citet{Huang:2006bp} are interested in the production of
$\gamma$-rays in an astrophysical environment, they have actually
published the differential cross sections for CR protons or
helium nuclei impinging on the ISM. In the former case, the
cross section is defined as
\beq
{\displaystyle \frac{d{\sigma}}{dE}}
\left( p + {\rm ISM} \right) \, = \, {\displaystyle \sum_{\rm A}} \;
X_{\rm A} \times {\displaystyle \frac{d{\sigma}}{dE}}
\left( p + {\rm A} \right) \;\; ,
\eeq
where the composition by number of the ISM has been set equal to
$X_{\rm H}  = 0.9$,
$X_{\rm He} = 0.1$,
$X_{\rm C} =  2 \times 10^{-4}$, and
$X_{\rm O} =  4 \times 10^{-4}$.
We assumed that composition to be constant throughout the
Milky Way gaseous component.

\vskip 0.1cm
The expression of the $\gamma$-ray flux at the Earth may be further
simplified into a convolution along the los of the hydrogen density
$n_{\rm H}$ with the effective emissivity ${\cal E}_{\rm eff}$ of
the ISM per hydrogen atom
\beq
\Phi_{\gamma}(l,b,E) \, = \, {\displaystyle \int}_{\rm \!\!\! los} ds \;
n_{\rm H} (\mathbf{x}) \, {\cal E}_{\rm eff} (\mathbf{x},E) \;\; .
\label{flux_nH_E_eff}
\eeq
This effective emissivity, which encodes the ISM composition and the
$\gamma$-ray production cross sections of the various nuclear channels,
is related to the \citet{Huang:2006bp} proton and helium cross sections
through
\beq
{\cal E}_{\rm eff}(\mathbf{x},E) \, = \,
{\displaystyle \frac{1}{X_{\rm H}}} \,
{\displaystyle \int}_{\!\!\! T_{\! min}}^{+ \infty} dT \;
\left\{
{\displaystyle \frac{d{\sigma}}{dE}}
\left(
p[T] + {\rm ISM} \to \gamma[E]
\right) \times \Phi_{p}(\mathbf{x},T)
\; + \;
{\displaystyle \frac{d{\sigma}}{dE}}
\left(
\alpha[T] + {\rm ISM} \to \gamma[E]
\right) \times \Phi_{\alpha}(\mathbf{x},T)
\right\} \;\; .
\label{def:emissivity_eff}
\eeq
The value of ${\cal E}_{\rm eff}$ at the solar system is readily obtained
by inserting the proton and helium fluxes observed at the
Earth in the righthand side of expression~(\ref{def:emissivity_eff}).
The red solid curve of Fig.~\ref{fig:local_emissivity} is based on the
\citet{2007APh....28..154S} measurements
and features $E^{2} {\cal E}_{\rm eff}$ as a function of photon energy
$E$. The agreement with the Fermi data \citep{Abdo:2009ka} is fairly
good, although the emissivity tends to be overpredicted.
This is slightly puzzling since room must be left for other processes
contributing to the $\gamma$-ray diffuse emission. Below 100 MeV,
inverse Compton scattering and bremsstrahlung take over pion production.
Although the former component has been substracted from the data, the
hadronic emissivity cannot exceed the total emissivity, and the red
solid line should be located below the Fermi points.

\vskip 0.1cm
The detailed investigation of that apparent discrepancy will be
presented elsewhere. Suffice it to say that the emissivity measured
by Fermi is not strictly local but is an average over the regions
of the Milky Way selected by \citet{Abdo:2009ka} for their analysis.
Those regions contain mostly atomic hydrogen and are located
farther away from the Galactic center than the solar circle is.
Although 85\% of the gas along the los lies within 1~kpc from the
Earth, the proton and helium fluxes to be used in
relation~(\ref{def:emissivity_eff}) are smaller than the values
borrowed from \citet{2007APh....28..154S}. How much smaller is
a crucial point on which depends the estimate of the bremsstrahlung
contribution to the local $\gamma$-ray diffuse radiation field.
Another possibility is to be found in the $\gamma$-ray production
cross section itself. The \citet{Huang:2006bp} parameterization could
overestimate photon production below a few tens of GeV, a region where
the \citet{Kamae:2006bf} model provides a robust alternative. The
uncertainty on the $\gamma$-ray emissivity induced by the existence
of different choices for the $\gamma$-ray production cross section will
be examined in section~\ref{sec:sigma}.

%
\subsection{Cosmic ray transport throughout the Galaxy}
\label{sec:CR_model}

%
%
The proton and helium fluxes are measured at the Earth but we need
to know them at any point $\mathbf{x}$ along the los. Deriving
$\Phi_{p}(\mathbf{x},T)$ and $\Phi_{\alpha}(\mathbf{x},T)$ from
their solar circle values is mandatory and requires that we model
the propagation of CR species inside the Galaxy.
Regardless of the mechanism responsible for their production, charged particles
propagate through the Galactic magnetic field and are deflected by its
irregularities: the Alfv\'en waves. In the regime where the magnetic turbulence
is strong -- which is the case for the Milky Way -- cosmic ray transport needs
to be investigated numerically. Monte Carlo simulations \citep{Casse:2001be}
indicate that it is similar to space diffusion with a coefficient
\beq
K(T) \, = \, K_{0} \; \beta \;
\left( {\mathcal R}/{\rm 1 \, GV} \right)^{\delta} \;\; ,
\label{space_diffusion_coefficient}
\eeq
which increases as a power law with the rigidity ${\mathcal R} = {p}/{q}$
of the nucleus.
In addition, because the scattering centers drift inside the Milky Way with
a velocity $V_{a} \sim$ 20 to 100 km s$^{-1}$, a second-order Fermi mechanism
is  responsible for some mild diffusive reacceleration. Its coefficient $K_{EE}$
depends on the cosmic ray velocity $\beta$ and total energy $E$ and is
related to the space diffusion coefficient $K(T)$ through
\beq
K_{EE} \, = \,
{\displaystyle \frac{2}{9}} \; V_{a}^{2} \;
{\displaystyle \frac{E^{2} \beta^{4}}{K(T)}} \;\; .
\eeq
Because $K_{EE}$ is inversely proportional to $K$, the typical timescale for
energy diffusion -- which is given by the ratio ${E^{2}}/{K_{EE}}$ -- is much
larger than the timescale associated to space diffusion  for energies
exceeding a few GeV per nucleon (GeV/n).
As we are interested in the $\gamma$-ray diffuse
emission in the GeV to TeV range, diffusive reacceleration can be safely
disregarded since photons are produced by protons and helium nuclei with
energies greater than 10 to 20~GeV/n.
The proton and helium fluxes with which we derive the
effective emissivity ${\cal E}_{\rm eff}$ at the solar circle in
sections~\ref{sec:sigma} and \ref{sec:phi_proton_alpha_at_earth} are simple
power laws obtained by fitting the various CR data . These power laws do
not exhibit any excess in the Gev region as we would have expected should
diffusive reacceleration be efficient at sub-GeV scales.
The $\gamma$-ray emissivity ${\cal E}_{\rm eff}({\odot})$ that we obtain, is
in fair agreement with the \citet{Abdo:2009ka} measurements down actually
to a photon energy of 100~MeV.
Energy losses do not play any major role either since they replenish
the low-energy regions of the CR proton and helium spectra where particles
are already abundant.
On the contrary, Galactic convection wipes cosmic rays away from the disk
with a velocity $V_{C} \sim$ 5 to 15 km s$^{-1}$. This process has some effect
at low energy.

%
\vskip 0.1cm
We assume that steady state holds for CR protons and helium nuclei.
The master equation fulfilled by the space and energy distribution function
$\psi = dn / dT$ of each of these CR species may be written as
\beq
\partial_{z} \left( V_{C} \, \psi \right)  \, - \, K  \, \Delta \psi
\, = \, q \left( {\mathbf x} , T \right) \;\; ,
\label{master_equation}
\eeq
where energy losses and diffusive reacceleration have been neglected.
%
%
This equation can be
solved within the framework of the semi-analytic two-zone model,
which has been extensively discussed in previous works such as
\citet{Maurin:2001sj} and \citet{Donato:2001ms}.
According to this approach, the region of the Galaxy inside which cosmic rays
diffuse -- the so-called diffusive halo or DH -- is pictured as a thick disk
that matches the circular structure of the Milk Way. The Galactic disk of stars
and gas, where primary cosmic rays are accelerated, lies in the middle. It extends
radially 20 kpc from the center and has a half thickness $h$ of 100~pc.
Confinement layers where cosmic rays are trapped by diffusion lie above and beneath
this thin disk of gas.
The intergalactic medium starts at the vertical boundaries $z = \pm L$, as well as
beyond a radius of $r = R_{\rm Gal} \equiv 20$ kpc. The half thickness
$L$ of the DH is still unknown, and reasonable values range from
1 to 15~kpc.
The diffusion coefficient $K$ is the same everywhere, whereas the convective velocity
is exclusively vertical with component $V_{C}(z) = V_{C} \; {\rm sign}(z)$. This Galactic
wind, which is produced by the bulk of the disk stars like the Sun, drifts away from
its progenitors along the vertical directions.
Also the normalization coefficient $K_{0}$, the index $\delta$, the
Galactic drift velocity $V_{C}$, and the Alfv\'en velocity $V_{a}$ are all unknown.
This situation can be remedied with the help of the boron-to-carbon ratio (B/C), which
is quite sensitive to cosmic ray transport and which may be used as a constraint.
The three propagation models featured in Table~\ref{table:prop} have been drawn
from \cite{2004PhRvD..69f3501D}. The MED configuration provides the best fit to the
B/C measurements, whereas the MIN and MAX models lead respectively to the minimal and
maximal allowed antiproton fluxes, which can be produced by the annihilation of
dark matter particles.

%
\vskip 0.1cm
The axial symmetry of the DH leads naturally to expand the CR density $\psi$
as a series of the Bessel functions $J_{0}({\alpha_{i}} \, {r}/{R_{\rm Gal}})$
where ${\alpha_{i}}$ is the i-th zero of the function $J_{0}$. This ensures that
$\psi$ vanishes at the radial boundary of the DH.
The source term accounts effectively for both the production ($q>0$) and the
destruction ($q<0$) of CR nuclei. The latter are accelerated by shock waves
driven by supernova explosions taking place in the Galactic disk. The CR nuclei
also undergo spallation reactions on the ISM, which depopulate the high-energy
tails of their spectra. In the thin disk approximation, the source term may be
expressed as
\beq
q \left( {\mathbf x} , T \right) \, = \, 2 \, h \, \delta_D(z) \;
\left\{ \,
\rho(r,z) \, Q_{\rm tot}(T) \; - \; \Gamma \, \psi
\, \right\} \;\; ,
\eeq
where \textbf{$\delta_D(z)$} is the Dirac function.
We have assumed that the energy and spatial dependencies of the primary CR
production rate can be disentangled. The space distribution $\rho(r,z)$ is
discussed in section~\ref{sec:f_SNR_radial}. It is normalized in such a way
that $Q_{\rm tot}(T)$ accounts for the total production of particles with
kinetic energy per nucleon $T$ over the whole Galaxy.
In the case of protons, the spallation term is given by
\beq
\Gamma_{\! p} \, = \, v_{p} \,
\left\{
\sigma_{\! p{\rm H}} \, n_{\rm H} \; + \; \sigma_{\! p{\rm He}} \, n_{\rm He}
\right\} \;\; ,
\eeq
where the densities $n_{\rm H}$ and $n_{\rm He}$ were averaged to $0.9$ and $0.1$ cm$^{-3}$ respectively. The total proton-proton cross section
$\sigma_{\! p{\rm H}}$ has been parameterized according to
\citet{Nakamura:2010zzi}, while $\sigma_{\! p{\rm He}}$ is related to
$\sigma_{\! p{\rm H}}$ by the \citet{Norbury:2006hp} \textbf{scaling factor $4^{(2.2)/3}$.}
The velocity of protons with kinetic energy $T$ is denoted by $v_{p}$.
%
%
The master equation~(\ref{master_equation}) simplifies considerably as long
as $z$ is different from $0$. This leads to the solution
\beq
\psi(r,z,T) \, = \,
{\displaystyle \sum_{i=1}^{\infty}} \; P_{i}(T) \times
\exp \left( {\displaystyle \frac{V_{C} \, |z|}{2 K}} \right) \times
\left\{
{\sinh \left[ {\displaystyle \frac{S_{i}}{2}} \left( L - |z| \right) \right]}
\, / \,
{\sinh \left[ {\displaystyle \frac{S_{i}}{2}}        L               \right]}
\right\} \times
J_{0}({\alpha_{i}} \, {r}/{R_{\rm Gal}}) \;\; ,
\label{equ:psi_bessel}
\eeq
where the inverse lengths $S_{i}$ are defined by
\beq
S_{i}(T) \, = \, \sqrt{\displaystyle
\left( {\displaystyle \frac{2 \alpha_{i}}{R_{\rm Gal}}} \right)^{2}
+
{\displaystyle \frac{V_{C}^{2}}{K^{2}}}} \;\; .
\eeq
Bessel expanding the primary CR radial distribution $\rho(r,0)$ and
vertically integrating the complete master equation~(\ref{master_equation})
through the thin disk allows, after some algebra, the Bessel
transforms $P_{i}$ to be expressed as
\beq
P_{i}(T) \, = \,
{\displaystyle \frac{q_{i}}{A_{i}}} \times Q_{\rm tot}(T) \;\; .
\eeq
The Bessel transforms $q_{i}$ of the primary CR source radial distribution
are related to $\rho(r,0)$ through
\beq
q_{i} \, = \,
{\displaystyle \frac{1}{J_{1}^{2}({\alpha_{i}})}} \times
{\displaystyle \frac{1}{\pi \, R_{\rm Gal}^{2}}}  \times
\left\{
{\displaystyle \int_{0}^{1}} u \, du \, J_{0}(\alpha_{i} u) \, \rho(r = u R_{\rm Gal},0)
\right\}
\, / \,
\left\{
{\displaystyle \int_{0}^{1}} u \, du \, \rho(r = u R_{\rm Gal},0)
\right\} \;\; ,
\eeq
and do not depend on the CR energy. The coefficients $A_{i}$ encapsulate
the effects of space diffusion and Galactic convection as well as the
spallations on the ISM. They are given by
\beq
A_{i}(T) \, = \,
K S_{i} \coth \left( {\displaystyle \frac{S_{i} L}{2}} \right) \; + \;
V_{C} \; + \;
2 \, h \, \Gamma \;\; .
\eeq
The total CR production rates $Q_{\rm tot}(T)$ are obtained by requiring
that the proton and helium fluxes at the Earth, which in our model are
defined as
\beq
\Phi_{p}({\odot} , T)      \, = \, {\displaystyle \frac{1}{4 \pi}} \, v_{p}
\, \psi_{p}({\odot} , T)
\;\;\;\; {\rm and} \;\;\;\;
\Phi_{\alpha}({\odot} , T) \, = \, {\displaystyle \frac{1}{4 \pi}} \, v_{\alpha}
\, \psi_{\alpha}({\odot} , T) \;\; ,
\eeq
are actually equal to the fluxes
$\Phi_{p}^{\rm exp}$ and $\Phi_{\alpha}^{\rm exp}$ measured by the various
CR experiments. In the case of protons, this translates into
\beq
Q_{{\rm tot},p}(T) \, = \,
{\displaystyle \frac{4 \pi}{v_{p}}} \times \Phi_{p}^{\rm exp}(T) \times
\left\{
{\displaystyle \sum_{i=1}^{\infty}} \;
{\displaystyle \frac{q_{i}}{A_{i}}} \;
J_{0}({\alpha_{i}} \, {r_{\odot}}/{R_{\rm Gal}})
\right\}^{-1} \;\; .
\eeq
The galactocentric distance of the Earth $r_{\odot}$ has been set equal to
8.5~kpc.
%
%
Our modeling allows us to \textit{retropropagate} CR protons and helium nuclei
from the Earth throughout the DH and to get
$\Phi_{p}(\mathbf{x},T)$ and $\Phi_{\alpha}(\mathbf{x},T)$ everywhere along
the los.

%
\begin{figure}[h!]
\centering
\includegraphics[width=\textwidth]{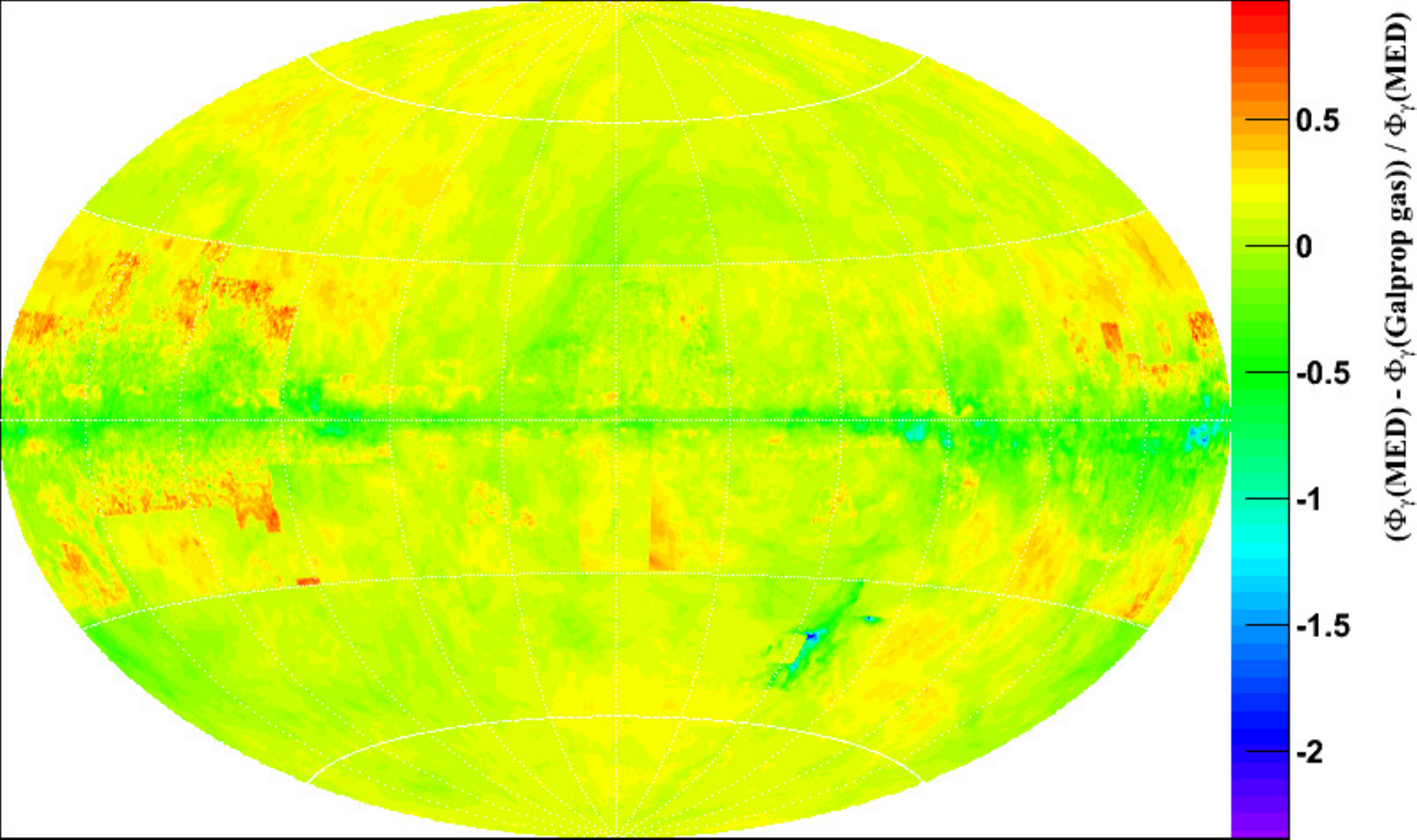}
\caption{
Our reference map of the diffuse $\gamma$-ray emission of the Milky Way
at 30 GeV obtained according to the method outlined in
section~\ref{sec:theory}. The dominant hadronic component alone is
considered. The cosmic ray proton and helium fluxes at the Earth's osition are
taken from \citet{2007APh....28..154S}. These fluxes were
\textit{retropropagated} throughout the DH with the MED model of
Table~\ref{table:prop}. The distribution of primary CR sources
in the Galactic disk was borrowed from \citet{2004IAUS..218..105L}.
The differential photon production cross sections of CR protons and
helium nuclei impinging on the ISM were parameterized according
to \citet{Huang:2006bp}. This map is based on the HI and CO 3D Galactic
distribution of \citet{2008ApJ...677..283P}. The $X_{\rm CO}$ factor was set equal to $2.3 \times 10^{20}$ molecules cm$^{-2}$ (K.km.s$^{-1}$)$^{-1}$ everywhere in the Galaxy.
}
\label{fig:ref_map}
%
\vskip -0.15cm %
%
\end{figure}

%
\subsection{The basic relations for a fast calculation}
\label{sec:basic_relations}

Equipped with the above notations, we are ready to derive the hadronic
component of the Galactic $\gamma$-ray diffuse emission. For a given photon
energy $E$, each map requires five nested summations. We first need to 
scan through the Galactic longitude $l$ and latitude $b$. Once a pixel
is chosen, the effective emissivity ${\cal E}_{\rm eff}$ of the ISM per
hydrogen atom is integrated along the los. This emissivity results 
from the convolution in energy of the differential photon production cross
sections with the proton and helium fluxes $\Phi_{p}(\mathbf{x},T)$ and
$\Phi_{\alpha}(\mathbf{x},T)$. The latter are finally obtained from the
expansion~(\ref{equ:psi_bessel}) over the corresponding Bessel coefficients
$P_{\! p,i}(T)$ and $P_{\! {\alpha},i}(T)$. If those five nested summations are
performed by brute force, each map requires a few hours of CPU time on an
ordinary PC.
That is why we have taken advantage of the Bessel formalism to speed up
the calculation. The idea is to never have more than three nested loops
to compute at any time and to decompose the calculation in a few imbricated
steps.

\vskip 0.1cm
We first computed the Bessel transforms $P_{\! p,i}(T)$ and
$P_{\! {\alpha},i}(T)$ from our preferred parameterization of the
proton $\Phi_{p}^{\rm exp}$ and helium $\Phi_{\alpha}^{\rm exp}$ fluxes
measured at the Earth's position -- our reference model is based on
\citet{2007APh....28..154S}. The kinetic energy per nucleon $T$ is
varied from $T_{\! min} = 0.28$~GeV up to $10^{3}$~TeV.
We then point out that the radial and vertical dependencies factor out
in the Bessel expansion~(\ref{equ:psi_bessel}). We may define
the intermediate integral over the kinetic energy per nucleon as
\bea
{\cal I}_{i}(z,E) & = &
{\displaystyle \frac{1}{X_{\rm H}}} \,
{\displaystyle \int}_{\!\!\! T_{\! min}}^{+ \infty} \! dT \;
{\displaystyle \frac{v_{\rm nuc}(T)}{4 \pi}} \times
\label{def:I_i_z} \\
& \times &
\left\{
{\displaystyle \frac{d{\sigma}}{dE}}
\left(
p[T] + {\rm ISM} \to \gamma[E]
\right) \times P_{\! p,i}(T) \times {\cal V}_{\! p,i}(T,z)
\; + \;
{\displaystyle \frac{d{\sigma}}{dE}}
\left(
\alpha[T] + {\rm ISM} \to \gamma[E]
\right) \times P_{\! {\alpha},i}(T) \times {\cal V}_{\! {\alpha},i}(T,z)
\right\} \;\; , \nonumber
\ena
which only depends on the Bessel order $i$ and on the height $z$, once the
$\gamma$-ray energy $E$ has been selected. The function ${\cal V}_{\! i}$
captures the vertical behavior of the i-th term in the Bessel
expansion~(\ref{equ:psi_bessel}) and is given by
\beq
{\cal V}_{\! i}(T,z) \, = \,
\exp \left( {\displaystyle \frac{V_{C} \, |z|}{2 K}} \right) \times
\left\{
{\sinh \left[ {\displaystyle \frac{S_{i}}{2}} \left( L - |z| \right) \right]}
\, / \,
{\sinh \left[ {\displaystyle \frac{S_{i}}{2}}        L               \right]}
\right\} \;\; .
\eeq
We finally map the effective emissivity everywhere in the DH by computing
the expansion
\beq
{\cal E}_{\rm eff}(r,z,E) \, = \,
{\displaystyle \sum_{i=1}^{\infty}} \;
J_{0}({\alpha_{i}} \, {r}/{R_{\rm Gal}}) \times
{\cal I}_{i}(z,E) \;\; ,
\label{def:E_r_z}
\eeq
up to a maximal Bessel order of 100.
Relation~(\ref{def:emissivity_eff}) is recovered by combining
equations~(\ref{def:I_i_z}) and (\ref{def:E_r_z}). The $\gamma$-ray flux
is then obtained by integrating along the los the hydrogen density $n_{\rm H}$
multiplied by the effective emissivity ${\cal E}_{\rm eff}$, which we
interpolate for any position $\mathbf{x}$ from its values on the $r$ and
$z$ grid of expression~(\ref{def:E_r_z}).
The entire process does not take more than a minute on an ordinary PC.

\vskip 0.1cm
We followed this procedure to compute the reference map
displayed in Fig.~\ref{fig:ref_map}. The photon energy $E$ was set
equal to 30~GeV. The proton and helium fluxes at the Earth's position were
parameterized according to \citet{2007APh....28..154S}. These fluxes
were \textit{retropropagated} throughout the DH with the MED set of
parameters of Table~\ref{table:prop}. The distribution of primary CR sources
in the Galactic disk was borrowed from \citet{2004IAUS..218..105L}.
The photon production differential cross sections of protons and helium nuclei
impinging on the ISM were taken from \citet{Huang:2006bp}. We used
the Galactic atomic and molecular hydrogen maps of \citet{2008ApJ...677..283P}.
These maps were derived from the
\citet{2005A&A...440..775K} HI and composite
\citet{2001ApJ...547..792D} CO surveys. The ionized hydrogen distribution
were parameterized according to the \citet{2002astro.ph..7156C} prescription.
The $X_{\rm CO}$ factor was set equal to $2.3 \times 10^{20}$ molecules cm$^{-2}$ (K.km.s$^{-1}$)$^{-1}$ everywhere in the Galaxy.
The main thrust of this analysis was to vary individually each of the many parameters involved in the problem and to calculate the variations in the $\gamma$-ray flux relatively to our reference model.

%
\section{Uncertainties arising from cosmic ray propagation}
\label{sec:cr_propagation_error}

As explained in the previous section, the propagation of CR particles in the Galaxy
is described by the four parameters $K_0$, $\delta$, $L$, and $V_C$ as long as diffusive
reacceleration is not taken into account. The values of these parameters are not
derived from first theoretical principles but are merely determined by a few ratios
of secondary to primary CR abundances. The B/C has been used
by \citet{Maurin:2001sj} to considerably reduce the parameter space.
Although more than 1,600 different sets of CR propagation parameters have been shown
in \citet{Maurin:2001sj} to be compatible with the B/C data, the three models featured
in Table~\ref{table:prop} provide a good indication of how uncertain Galactic CR
propagation is. The extreme models MIN and MAX gauge the possible spread
of values within which the various parameters are to be found. As shown in \citet{2004PhRvD..69f3501D}, these propagation
models correspond respectively to minimal, medium, and maximal fluxes of primary
antiprotons whose exotic dark matter species would produce everywhere inside the DH, if they exist. The more recent analysis
by \citet{2010A&A...516A..66P} makes use of Monte Carlo techniques and gives similar
results. That analysis confirms in particular that the  half thickness $L$ of the DH
is not determined and may vary between 1 and 16~kpc.
This also agrees, although to a somewhat lesser extent, with the conclusion reached by~\cite{2011ApJ...729..106T}, who find that $L$ may extend from 2 to 10~kpc.
The uncertainties on CR propagation translate into uncertainties on the
$\gamma$-ray flux. The spatial distributions of protons $\Phi_{p}(\mathbf{x},T)$
and helium nuclei $\Phi_{\alpha}(\mathbf{x},T)$ are modified if the
propagation parameters are varied. If the DH itself is expanded and high values
of $L$ are selected, CR species may illuminate hydrogen far away from the Galactic
disk, and the intensity of the $\gamma$-ray diffuse emission can be drastically
increased.
%
\begin{table}[h!]
\begin{center}
\begin{tabular}{|c||c|c|c|c|c|c|}
\hline
Model  & $\delta$ & $K_0$ [kpc$^2$/Myr] & $L$ [kpc] & $V_{C}$ [km/s] & $V_{a}$ [km/s] \\
\hline \hline
MIN  & 0.85 &  0.0016 & 1  & 13.5 &  22.4 \\
MED  & 0.70 &  0.0112 & 4  & 12   &  52.9 \\
MAX  & 0.46 &  0.0765 & 15 &  5   & 117.6 \\
\hline
\end{tabular}
\end{center}
\caption{
Typical combinations of CR propagation parameters that are compatible with the B/C
analysis by \citet{Maurin:2001sj}.}
\label{table:prop}
%
%
\end{table}
%

%
\begin{figure*}[h!]
\centering
\includegraphics[width=\textwidth]{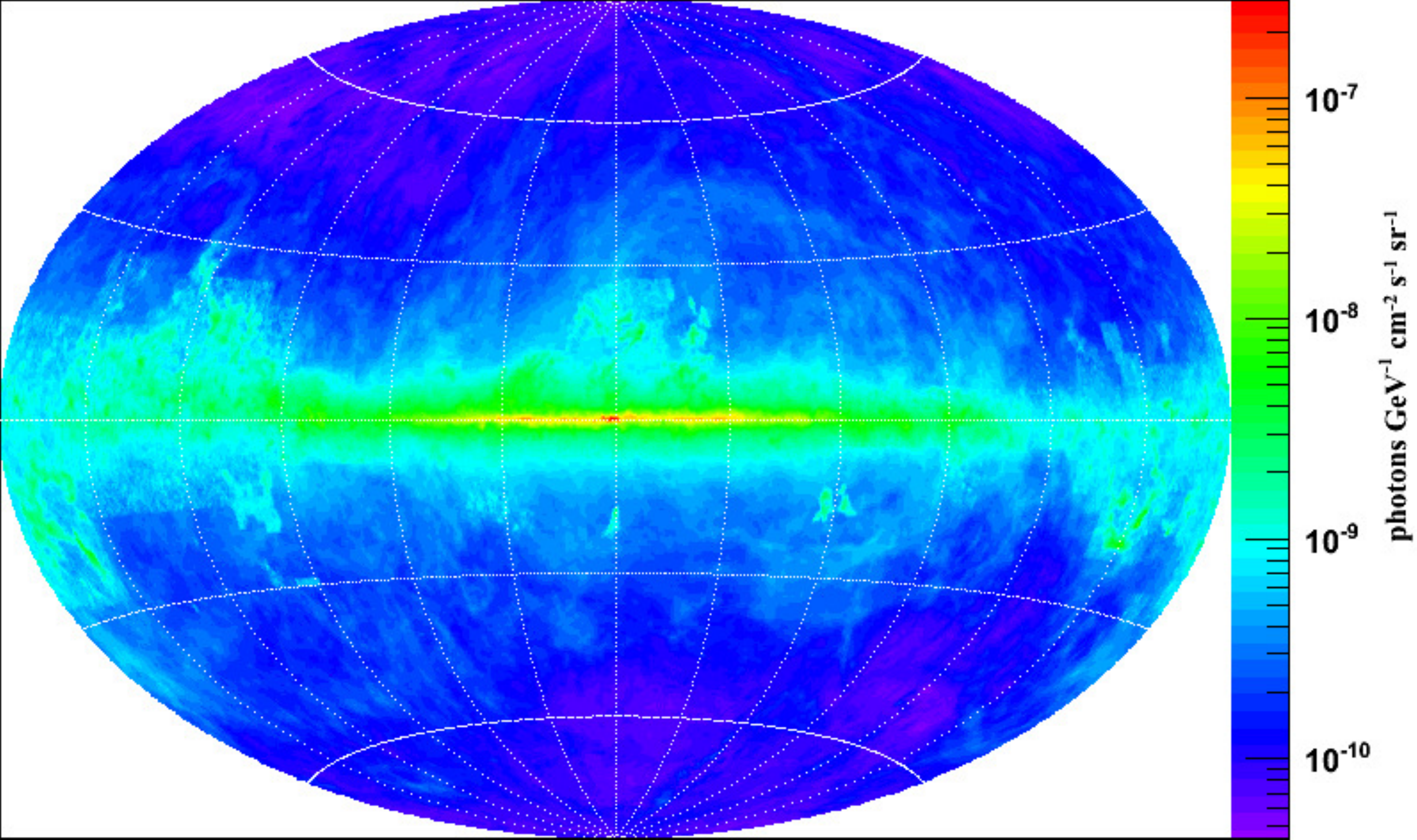}
\caption{
This sky map features the difference between the MIN and MAX models relative
to the MIN model. For each pixel we plotted the ratio
${\rm (map1 - map2) / map1}$
where map1 (map2) has been derived exactly like the $\gamma$-ray reference map
of Fig.~\ref{fig:ref_map} with the sole difference of using the MIN (MAX)
propagation parameters instead of the MED model.
}
\label{fig:minMax}
\end{figure*}

\vskip 0.1cm
To get a sense of how important CR propagation is for the Galactic
$\gamma$-ray diffuse emission, we plotted in Fig.~\ref{fig:minMax} the
difference between the MIN and MAX models relative to the MIN model. We 
first derived the $\gamma$-ray sky maps corresponding to the MIN and MAX sets
of propagation parameters of Table~\ref{table:prop} while keeping all the other
inputs of the reference map of Fig.~\ref{fig:ref_map} unchanged -- the latter
is based on the MED model. We then computed the contrast
\beq
{\cal C} \, = \,
\left(
{\Phi_{\gamma}({\rm MIN}) - \Phi_{\gamma}({\rm MAX})}
\right) /
{\Phi_{\gamma}({\rm MIN})} \;\; ,
\eeq
and plotted its value for each of the pixels of the map displayed in
Fig.~\ref{fig:minMax}. As this map shows clearly, the uncertainty due to CR
propagation can be quite important. For a photon energy of 30~GeV, the ratio of
the MAX to MIN $\gamma$-ray fluxes varies over the sky from
0.63 to 2.26. The former value corresponds to a contrast ${\cal C}$ of
+37\% and to the red central region of the Galactic disk, while the
latter value is associated to a contrast of -126\% and to a few dark blue spots
located at fairly low Galactic latitude. We found that the contrast
does not change much with energy in the GeV-TeV range. The map features
four distinct colored regions.

\vskip 0.1cm
\noindent
{\bf (i)}
The contrast vanishes inside the yellow domains. These domains correspond
to directions along which the gas is mostly nearby. The proton and helium
fluxes are then approximately given by their values at the Earth. Whatever
the CR model used to \textit{retropropagate} primary CR fluxes, we do not
expect them to vary much in the neighborhood of the solar system, hence
identical $\gamma$-ray emissivities and fluxes for the MIN and MAX
configurations.

\vskip 0.1cm
\noindent
{\bf (ii)}
In the green regions, the contrast lies between -50\% and -20\% as indicated
by the color scale on the righthand side of the map. The MAX $\gamma$-ray
flux is 20\% to 50\% larger than the MIN one. This moderate difference
can be interpreted if most of the gas that contributes to the signal is
close to the Galactic plane. For
the sake of the argument, we disregard all CR transport mechanisms except
space diffusion and assume that the DH has no radial boundaries, which
amounts to seting $R_{\rm gal}$ at infinity. We also concentrate on the
1D problem with CR densities depending only on the height $z$. The master
equation~(\ref{master_equation}) then yields the very simple solution
\beq
{\displaystyle \frac{\Phi_{p}(\mathbf{x},T)}{\Phi_{p}(\odot,T)}} \, = \,
{\displaystyle \frac{\Phi_{\alpha}(\mathbf{x},T)}{\Phi_{\alpha}(\odot,T)}} \, = \,
1 \, - \, {\displaystyle \frac{z}{L}} \;\; ,
\eeq
where $L$ is the half thickness of the DH. Values for that parameter range
form $L_{1} = 1$~kpc for the MIN model up to $L_{2} = 15$~kpc in the MAX
case. The $\gamma$-ray effective emissivity ${\cal E}_{\rm eff}$ also
scales as $1 - z/L$.
If the fine-grained distribution of gas $n_{\rm H} (\mathbf{x})$
that enters in relation~(\ref{flux_nH_E_eff}) is now replaced
by a homogeneous slab distribution extending from
$z = - L_{\rm H}$ to $z = + L_{\rm H}$, the expression for the
$\gamma$-ray flux boils down to
\beq
\Phi_{\gamma} \propto L_{\rm H} \, - \,
{\displaystyle \frac{L_{\rm H}^{2}}{2 L}} \;\; ,
\eeq
provided that $L_{\rm H} \leq L_{1}$. The contrast is given by
\beq
- \, {\cal C} \, = \, {\displaystyle
\frac{1 \, - \, ({L_{1}}/{L_{2}})}{({2}{L_{1}}/{L_{\rm H}}) \, - \, 1}}
\;\; .
\label{equ:simple_scaling_LH}
\eeq
A contrast of -50\% translates into a gas half thickness
$L_{\rm H} = 0.7$~kpc while a value of -20\% implies 
$L_{\rm H} = 0.35$~kpc. The green domains correspond to directions
for which most of the gas is within 0.35 to 0.7~kpc from the Galactic plane.

%
\begin{figure}[h!]
\centering
\includegraphics[width=0.6\columnwidth]{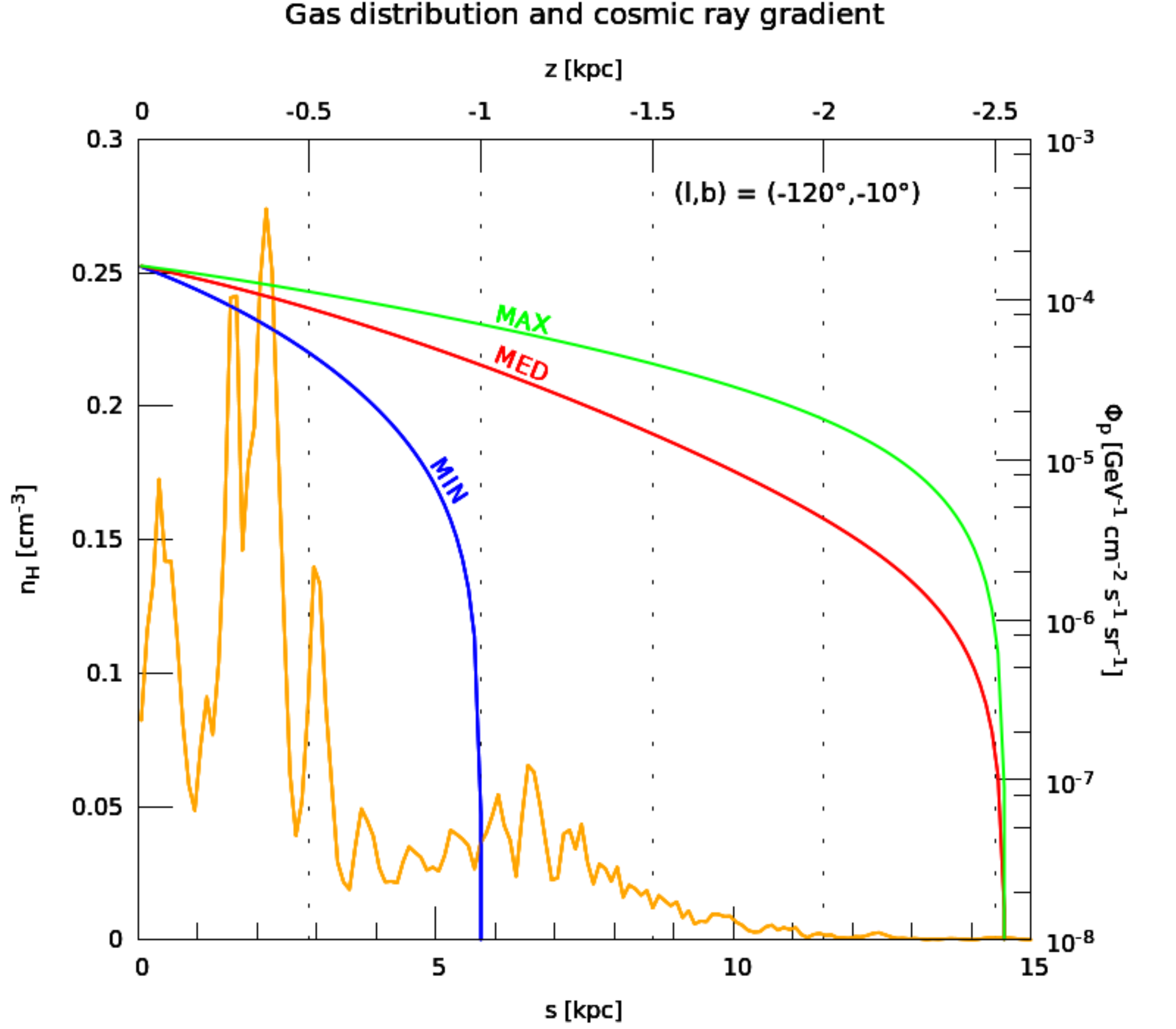}
\caption{
The gas density (orange with units on the left $y$-axis) and the CR proton flux
$\Phi_{p}(\mathbf{x})$ (units on the right $y$-axis) are plotted along the
los in the direction specified by the Galactic longitude and latitude
$(l , b) = (-120^{\circ} , -10^{\circ})$. The three models of
Table~\ref{table:prop} have been considered for the calculation of
the proton flux. They lead to the blue (MIN), red (MED) and green (MAX)
curves. As an illustration, the CR proton total energy $E_{p} = m_{p} + T$
has been set equal to 30~GeV. The relative distribution of primary nuclei
along the los is not expected to change much with energy.
}
\label{fig:minMaxwhy}
%
\vskip -0.25cm %
%
\end{figure}

\vskip 0.1cm
\noindent
{\bf (iii)}
Some regions of the sky seem more affected by CR propagation than others.
The dark blue spots of Fig.~\ref{fig:minMax} have a strong contrast with
a MAX flux more than twice as large as the MIN one.
Relation~(\ref{equ:simple_scaling_LH}) indicates that $L_{\rm H}$
overcomes $L_{1}$ when the contrast is less than -94\%.
Crudely speaking, the dark blue spots correspond to molecular clouds
located farther away than 1~kpc from the Galactic plane. These clouds
cannot be illuminated by CR protons and $\alpha$ particles in the MIN
configuration because they are outside the DH. The density $\psi$ of
any CR species indeed becomes rapidly negligible near the boundary of
the DH and vanishes completely outside.
A careful inspection is necessary though. As is clearly illustrated in
Fig.~\ref{fig:minMaxwhy}, the gas density, featured by the orange curve
with units on the left vertical axis, is not
at all negligible between $z = - 0.7$~kpc and $z = - 1.7$~kpc in the
direction of the large spot beneath the Galactic plane at
$(l , b) = (-120^{\circ} , -10^{\circ})$.
In the case of the MIN configuration, which corresponds to the blue line
with units on the right vertical axis, CR protons cannot interact much
with the gas located beyond a distance $s$ on the los of 4~kpc. Behind
that point, the gas clouds are too close to the vertical boundary of
the DH to be strongly illuminated. For a distance $s$ exceeding 6~kpc,
the clouds are even outside the DH as their height $z$ becomes less
than $- L_{1}$.
In the MED (red) and MAX (green) cases, the gas lying between
$z = - 0.7$~kpc and $z = - 1.7$~kpc is, on the contrary, fully contained
within the DH, hence a much larger $\gamma$-ray flux owing to a larger
abundance of hydrogen producing it.
We conclude that the blue spots correspond to clouds
located typically beyond a distance to the Galactic plane of 0.7~kpc
and lying therefore in a region where the MIN CR flux decreases 
rapidly, however, this effect is exaggerated by our sharp boundary
conditions. In a model where the space diffusion coefficient $K$ would
vary with $z$ like in \citet{2010PhRvD..82d3505P}, the relative increase
from the MIN to the MAX cases would not exceed a factor of 2.

\vskip 0.1cm
\noindent
{\bf (iv)}
Finally, the contrast ${\cal C}$ becomes positive in the central
red region of the Galactic plane where $|l| \leq 60^{\circ}$. In this part
of the sky, the $\gamma$-ray flux derived from the MIN propagation set
becomes larger than the flux obtained with the MAX model. This puzzling
behavior is not surprising. The MIN set of parameters is associated to
a strong convection and a weak space diffusion compared to the MAX case.
The primary CR nuclei are blown away efficiently from the Galactic disk.
They hardly make it to the Earth as they travel from the molecular ring,
located at 4~kpc from the Galactic center, where most of the primary CR
sources are found. In section~\ref{sec:CR_model}, we have derived the proton
and helium densities everywhere inside the DH by \textit{retropropagating}
these species back from the Earth where their fluxes are set equal to the
values provided by observations. If convection in the MIN case prevents a
substantial fraction of CR primaries to propagate from their sources to the
Earth, the total Galactic production rates $Q_{\rm tot}$ need to be enhanced
with respect to the MAX configuration in order to get the same proton and
helium abundances at the solar circle. The sources need to be
strengthened in the MIN case in order to give the observed proton and
$\alpha$ fluxes at the Earth. The $\gamma$-rays are a unique probe of the
CR densities inside the DH and not at the Earth only. Brighter CR sources
in the MIN case translate into a more powerful illumination of the ISM in
the direction of these sources. That is why the $\gamma$-ray diffuse
emission from the inner parts of the Galactic disk is the brightest in
the MIN model.

%
\begin{figure}[h!]
\centering
\includegraphics[width=0.60\textwidth]{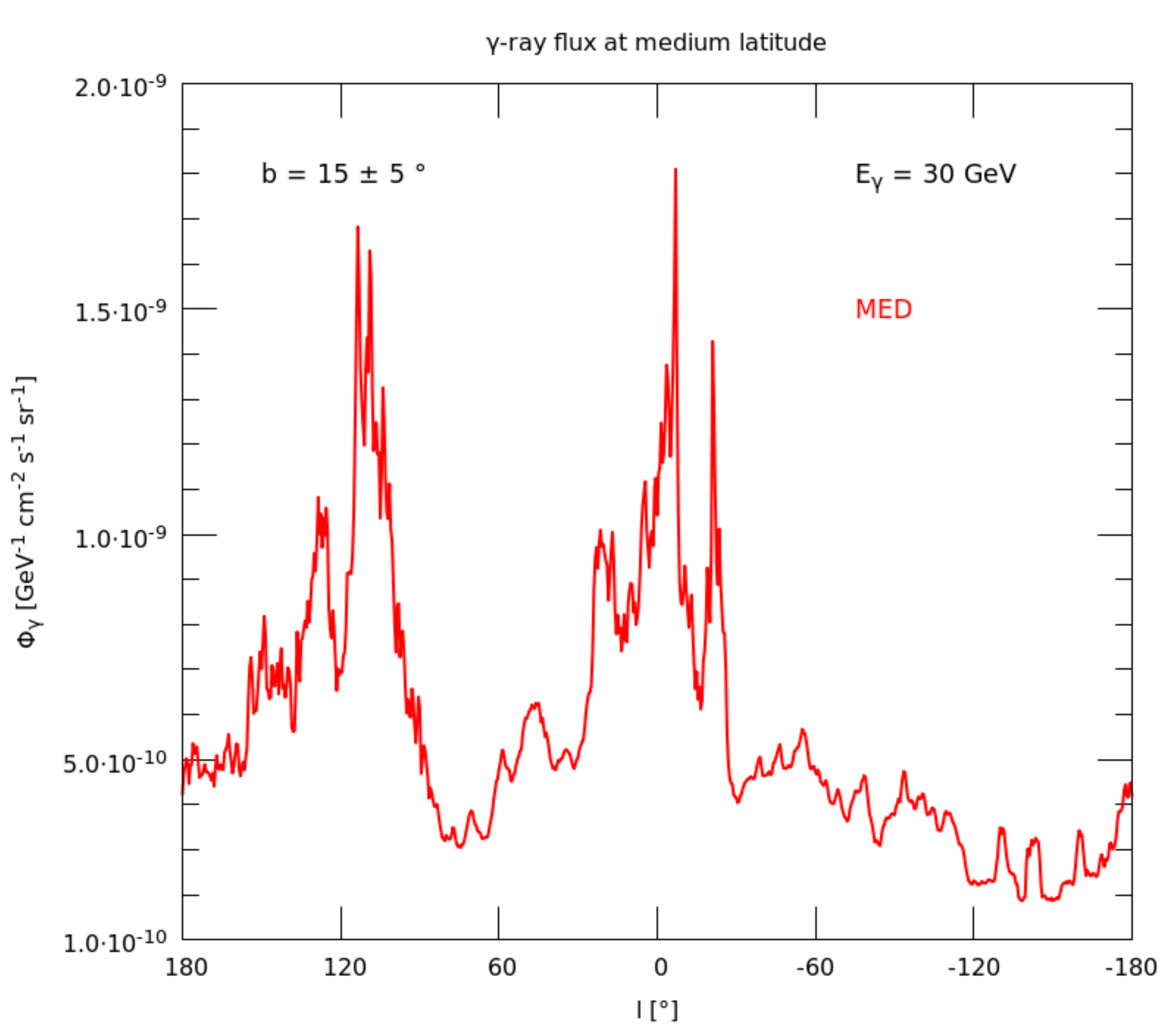}
\caption{
The $\gamma$-ray flux of the reference map of Fig.~\ref{fig:ref_map}
has been averaged over the Galactic latitude
$10^{\circ} \leq b \leq 20^{\circ}$ and plotted as a function of
Galactic longitude $l$. The photon energy is equal to 30~GeV.
The Galactic CR propagation parameters correspond to the MED model
of Table~\ref{table:prop}.
}
\label{fig:b15_E30}
%
\vskip -0.25cm %
%
\end{figure}

\vskip 0.1cm
However, not all parameters have a priori the same impact on the $\gamma$-ray
diffuse emission. That is why we systematically explored the effect of each
of them. We first extracted the band $10^{\circ} \leq b \leq 20^{\circ}$
from the reference map of Fig.~\ref{fig:ref_map}. That region has been
studied in some detail by the Fermi collaboration \citep{2009PhRvL.103y1101A}.
We then averaged the $\gamma$-ray flux over Galactic latitude $b$
and plotted the result in Fig.~\ref{fig:b15_E30} as a function of Galactic
longitude $l$. As in Fig.~\ref{fig:ref_map}, the photon energy
is 30~GeV. The flux varies up to a factor of four from one longitude $l$ to
another. The gas distribution is far from homogeneous and the two peaks at
$l = 0^{\circ}$ and $l = 120^{\circ}$ correspond to high concentrations
of hydrogen along the los.

%
\begin{figure*}[h!]
\centering
\includegraphics[width=0.45\textwidth]{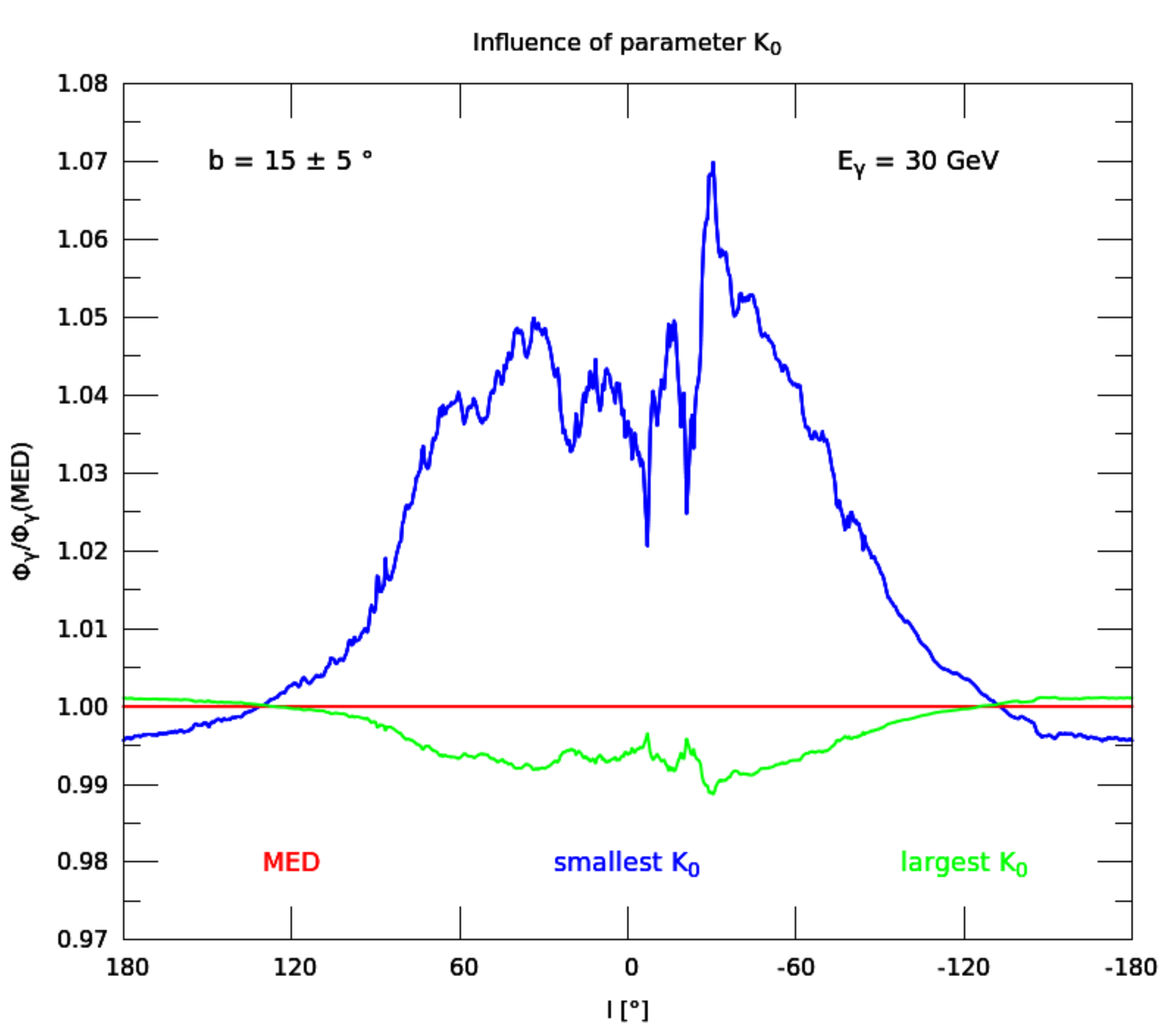}
\includegraphics[width=0.45\textwidth]{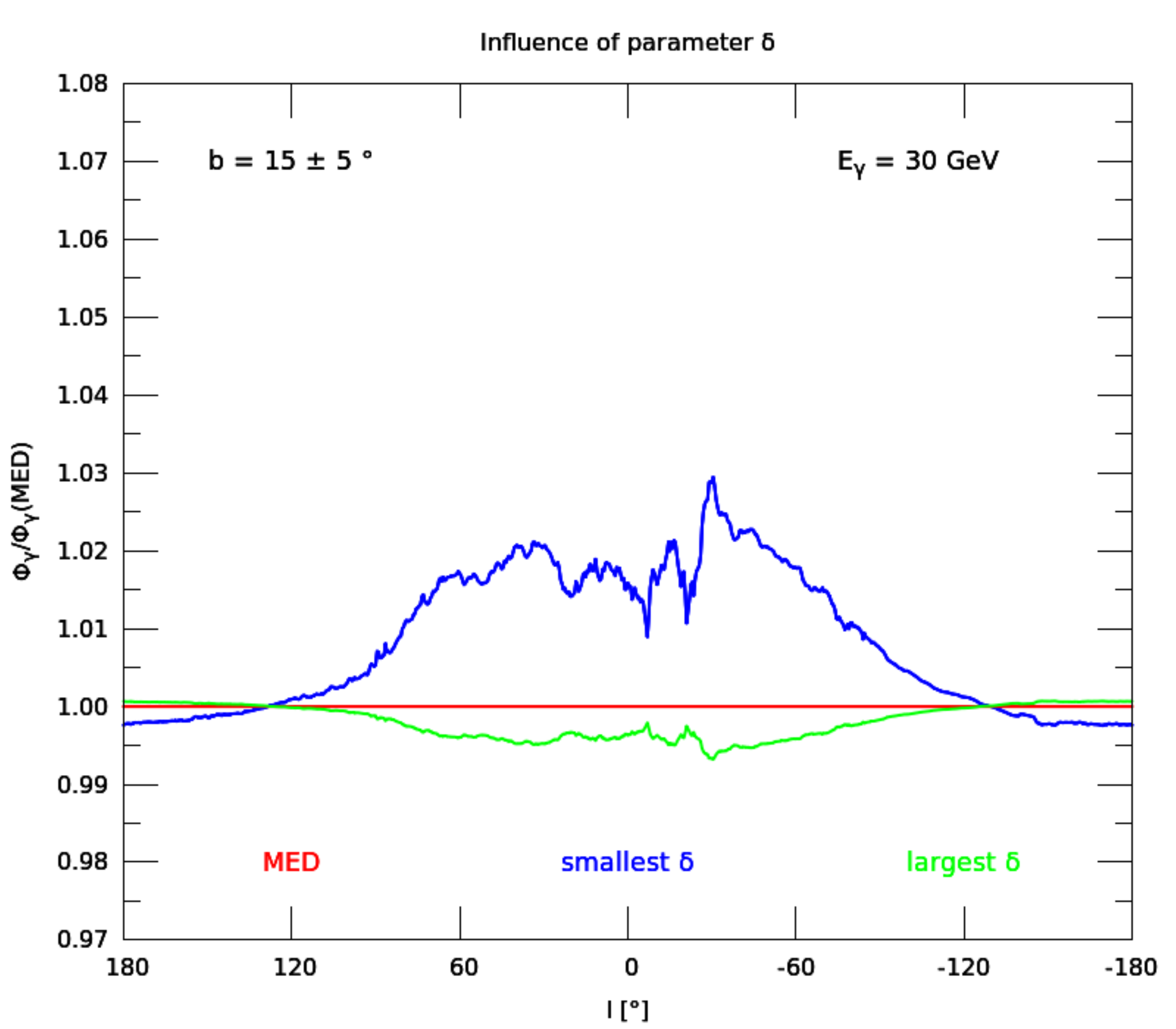}
\caption{
Variations in the $\gamma$-ray flux relative to the reference case of
Fig.~\ref{fig:b15_E30}. The red line is obtained when all the propagation
parameters are extracted from the MED set of Table~\ref{table:prop}.
\textit{Left panel:} changing only the normalization $K_{0}$ of the space
diffusion coefficient from its MED value to the lowest
($1.6  \times 10^{-3}$~kpc$^2$/Myr) and highest
($7.65 \times 10^{-2}$~kpc$^2$/Myr) values allowed by the B/C analysis yields
the blue and green curves respectively.
\textit{Right panel:} same as in the left panel but varying the spectral index
$\delta$ instead of the normalization $K_{0}$. The blue and green curves
are derived from the smallest (0.46) and largest (0.85) values
of the spectral index respectively.
}
\label{fig:K_et_D_rel}
%
\vskip -0.25cm %
%
\end{figure*}

\vskip 0.1cm
As featured in the two panels of Fig.~\ref{fig:K_et_D_rel}, varying the normalization
$K_{0}$ or the spectral index $\delta$ of the space diffusion coefficient $K$ has
little impact on the $\gamma$-ray diffuse emission. At that latitude and for that
energy, the variation reaches at most 7\% and 3\%. The strongest effect
is obtained when either $K_{0}$ or $\delta$ is decreased. As space diffusion becomes
less efficient, Galactic convection takes over, and the above-mentioned argument
applies. More primary CR nuclei are washed out of the DH as they propagate from
their production site, located at the molecular ring, to the Earth. The CR sources
need to be slightly brighter than in the MED configuration in order to maintain
the proton and helium fluxes at their terrestrial measured values, hence a larger
$\gamma$-ray flux towards the Galactic center. For the same reason, the
densities of CR protons and helium nuclei drop faster than in the MED case at large
galactocentric distances, in the outer fringes of the DH. We anticipate a small
drop in the $\gamma$-ray flux in the direction of the Galactic anti-center. This
is exactly what the blue curves illustrate in each of the panels of
Fig.~\ref{fig:K_et_D_rel}. Those curves are above the reference red line for a
Galactic longitude in the range
$- 120^{\circ} \leq l \leq 120^{\circ}$, while they exhibit a small deficit
towards the Galactic anti-center. The green curves feature the opposite behavior.
Space diffusion becomes more efficient than in the MED case, and the convection
argument can be reversed with slightly fainter CR sources and a small deficit of
the $\gamma$-ray flux at small Galactic longitudes.

%
\begin{figure*}[h!]
\centering
\includegraphics[width=0.45\textwidth]{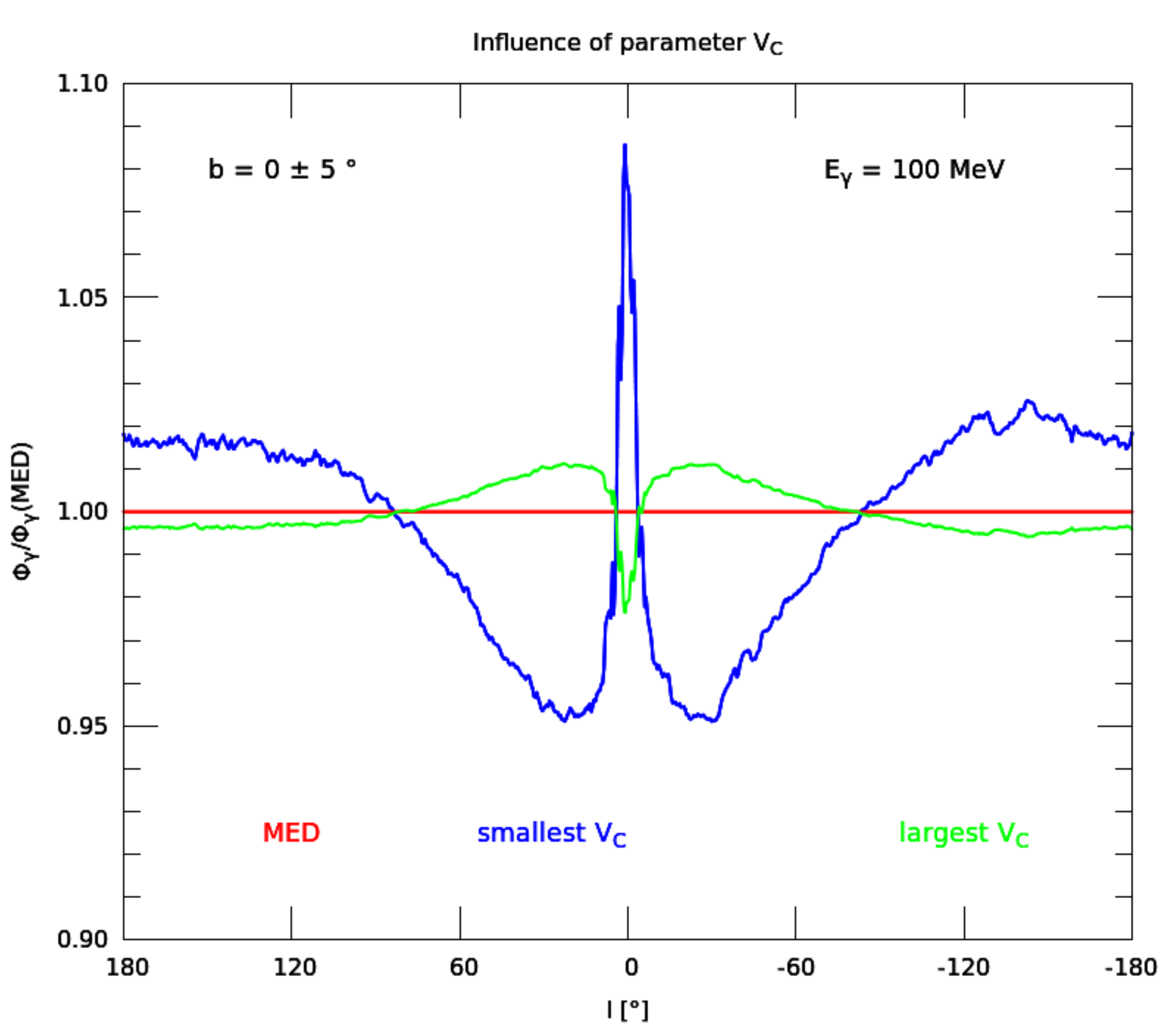}
\includegraphics[width=0.45\textwidth]{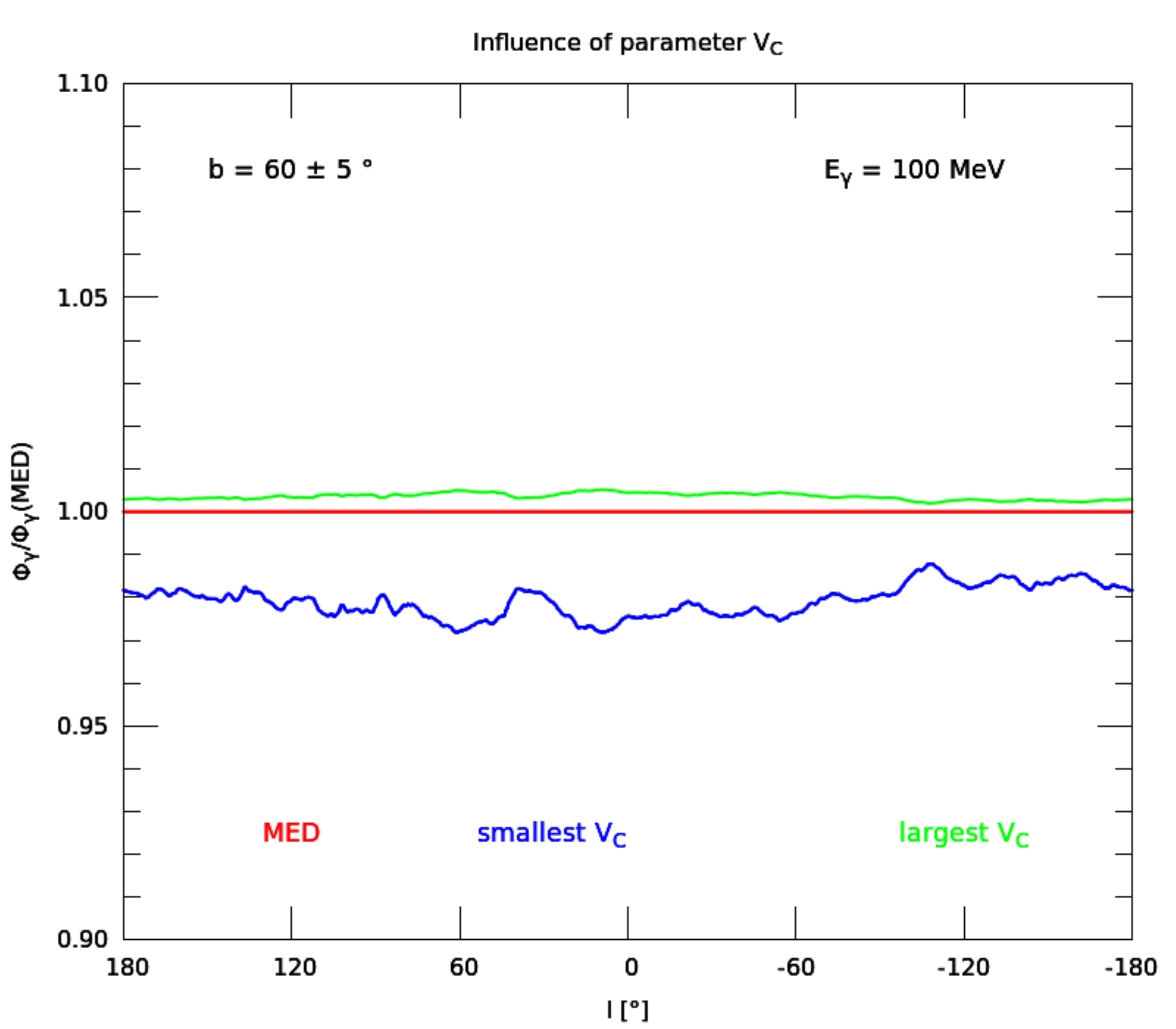}
\caption{
Same as in Fig.~\ref{fig:K_et_D_rel} but varying the convective wind
velocity $V_{C}$ instead of the normalization $K_{0}$ or the spectral
index $\delta$ of the space diffusion coefficient.
The blue and green curves correspond to the lowest
(5 km/s) and highest (14 km/s) values of $V_{C}$ respectively.
While fairly small at 30~GeV (left panel), the effect increases
at 1~GeV (right panel).
}
\label{fig:VC_rel}
%
\vskip -0.25cm %
%
\end{figure*}

\vskip 0.1cm
If Galactic convection is now amplified with respect to the MED reference
case while keeping all the other parameters fixed, we would expect the same
effect as if space diffusion were inhibited. According to our reasoning based
on the competition between convection and diffusion, an increase of $V_{C}$
forces the Galactic primary CR production rates $Q_{\rm tot}$ to be higher
in order to maintain the proton and helium fluxes at the Earth's position at their
measured values. Convection prevents CR particles from propagating along the
Galactic disk and wipes them out along the vertical directions of the DH. The
molecular ring becomes brighter, and the $\gamma$-ray emission from that
region is enhanced. The green curves of Fig.~\ref{fig:VC_rel} correspond
to a Galactic wind of 14~km/s and are located above the MED red lines. The
increase is quite small insofar as $V_{C}$ has just been slightly increased
from its MED value of 12~km/s. If $V_{C}$ is now decreased down to 5~km/s,
we get the blue curves located below the reference red lines. The effect is
more significant, although it does not exceed 0.5\% at 30~GeV (left panel).
At lower $\gamma$-ray energies, the CR proton and $\alpha$ particles implied
in the photon production also have lower energies and diffuse less efficiently
throughout the Milky Way magnetic field. Convection becomes relatively more
important with respect to diffusion. The same variations of $V_{C}$ are then
expected to have a stronger impact on the $\gamma$-ray flux. This is exactly
what the right panel of Fig.~\ref{fig:VC_rel} illustrates. The photon
energy has been set equal to 1~GeV. The blue line indicates a maximal
variation of 5\% at $l = - 30^{\circ}$, one order of magnitude more
than in the left panel where $E = 30$~GeV.
Finally the various curves of Figs.~\ref{fig:K_et_D_rel} and
\ref{fig:VC_rel} all have the same shape. In particular, the blue and green
lines are symmetrically drawn around the red axes.

%
\begin{figure}[h!]
\centering
\includegraphics[width=0.45\textwidth]{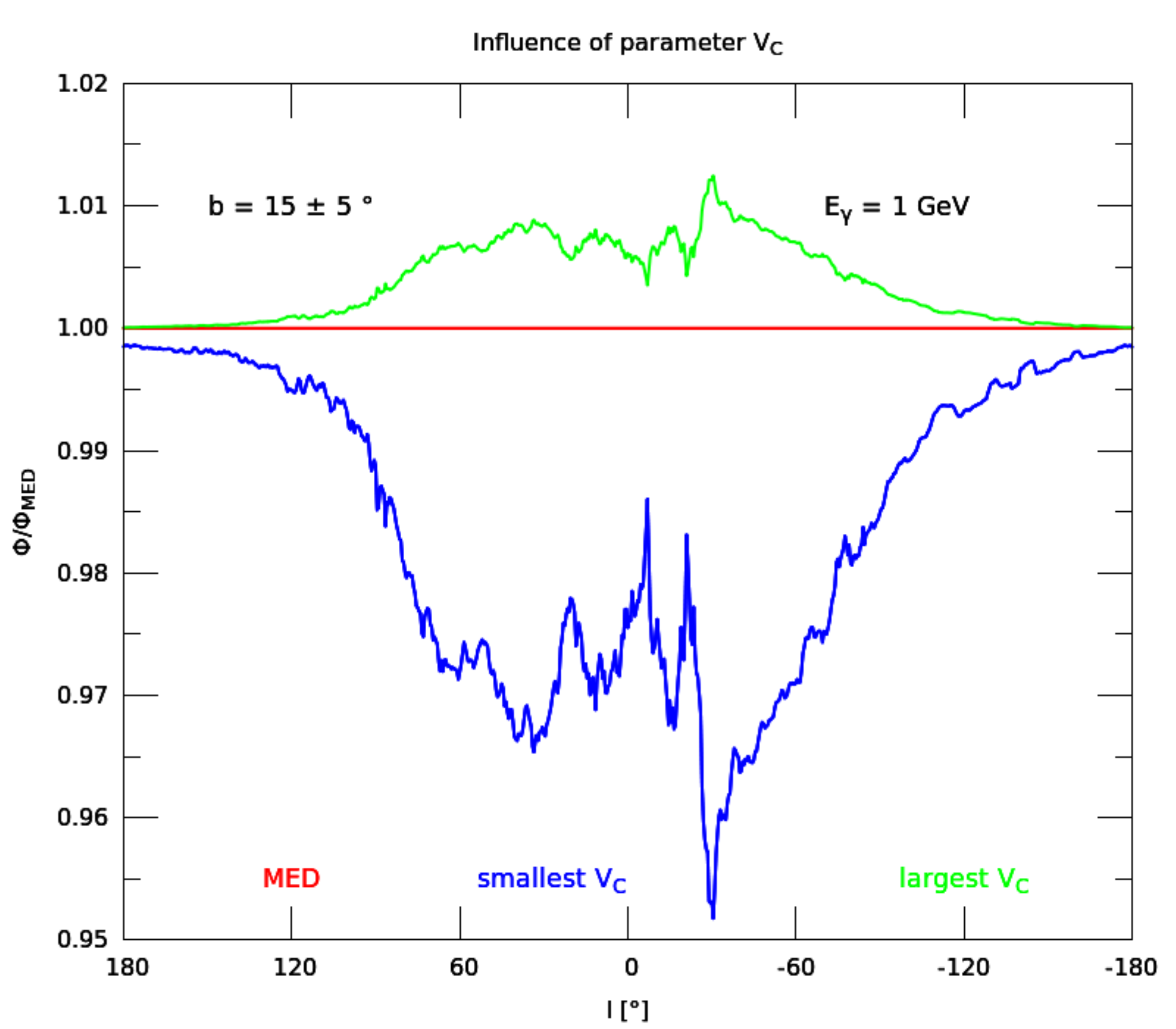}
\includegraphics[width=0.45\textwidth]{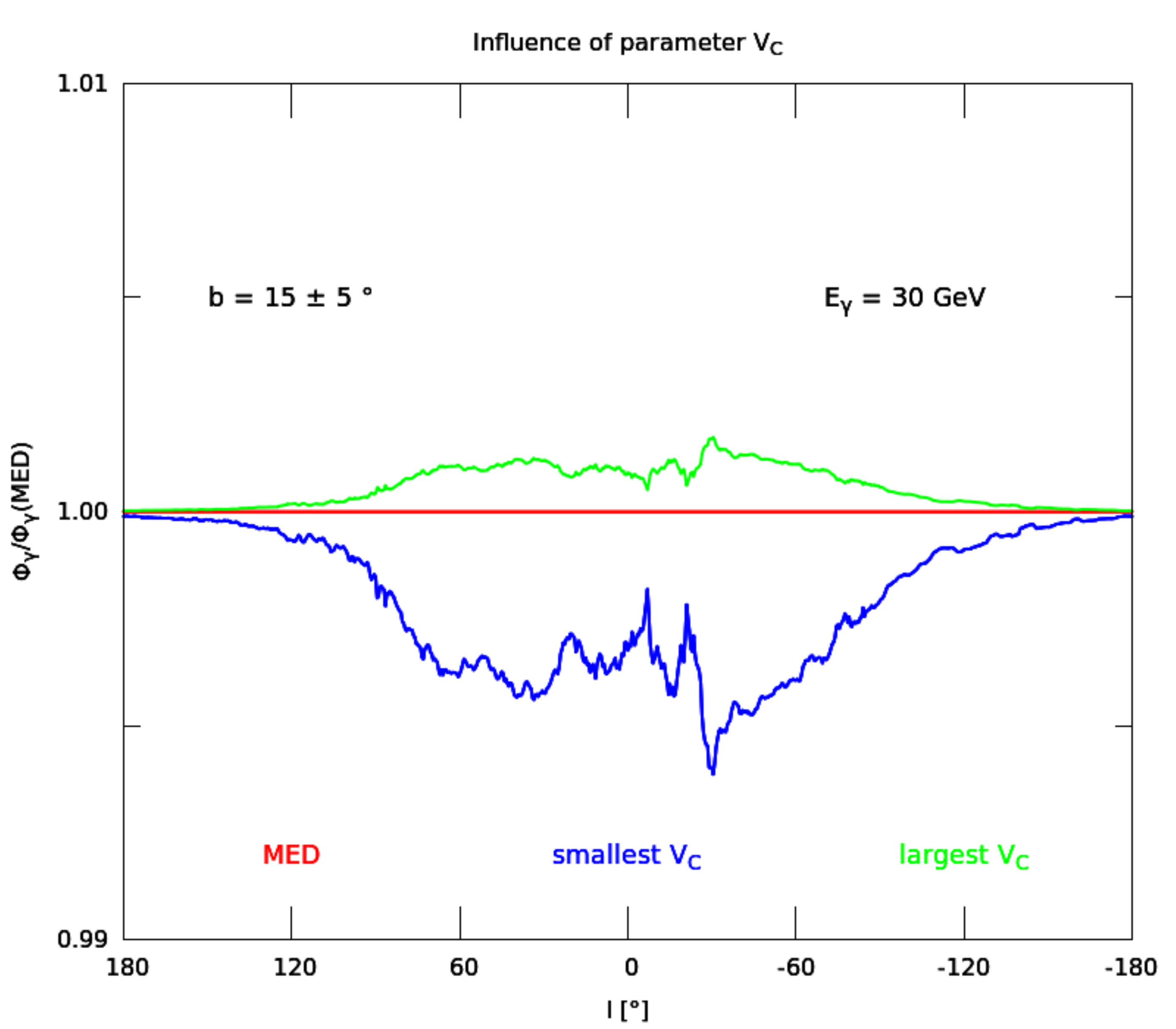}
\caption{
Same as in Fig.~\ref{fig:VC_rel} but for a $\gamma$-ray energy of 100~MeV
and at latitudes of $0^{\circ}$ (left panel) and $60^{\circ}$ (right panel).
The effect of convection is greatest along the Galactic plane and
fades as the height $|z|$ increases.
}
\label{fig:VC_rel_E_et_b}
%
\vskip -0.25cm %
%
\end{figure}

\vskip 0.1cm
Convection has an even greater impact at 100~MeV as shown in the left panel
of Fig.~\ref{fig:VC_rel_E_et_b} where the Galactic latitude $b$ has been set
equal to $0^{\circ} \pm 5^{\circ}$. This plot displays the variations in the
$\gamma$-ray diffuse emission along the Galactic plane as the wind velocity
$V_{C}$ is increased up to 14~km/s or decreased down to 5~km/s.
In the former case and for reasons already discussed above, the molecular ring
becomes clearly visible. The green curve actually overcomes the red axis for a
longitude $5^{\circ} \leq |l| \leq 85^{\circ}$, and the deficit in the
direction of the Galactic center at $-5^{\circ} \leq l \leq 5^{\circ}$. The CR
proton and helium nuclei located inside the molecular ring, at galactocentric
distances less than $\sim$~4~kpc, are not as abundant as in the MED reference
situation. The enhanced convection blows them away as they diffuse inwards
from the molecular ring so that the Galactic center is underpopulated compared to the MED case.
The blue line features the opposite behavior with a marked peak at the
Galactic center where CR primaries are now more abundant. The overall effect
does not exceed 9\%, however.
The Galactic plane is more sensitive to convection than
high latitude regions where diffusion tends to homogenize cosmic rays. In
the right panel of Fig.~\ref{fig:VC_rel_E_et_b}, the latitude has been set
equal to $60^{\circ} \pm 5^{\circ}$. The green (blue) curve is still above
(below) the red reference axis as a signature of the higher (lower) primary
proton and helium nuclei production rates $Q_{\rm tot}$. The variations
in the $\gamma$-ray flux do not exceed 3\%.
We finally decreased the photon energy down to 10~MeV, well below the
energy range to which this analysis is devoted. At these very low energies,
the whole sky becomes equally sensitive to the value of $V_{C}$ but the
effect remains less than 10\%.

%
\begin{figure*}[h!]
\centering
\includegraphics[width=0.60\textwidth]{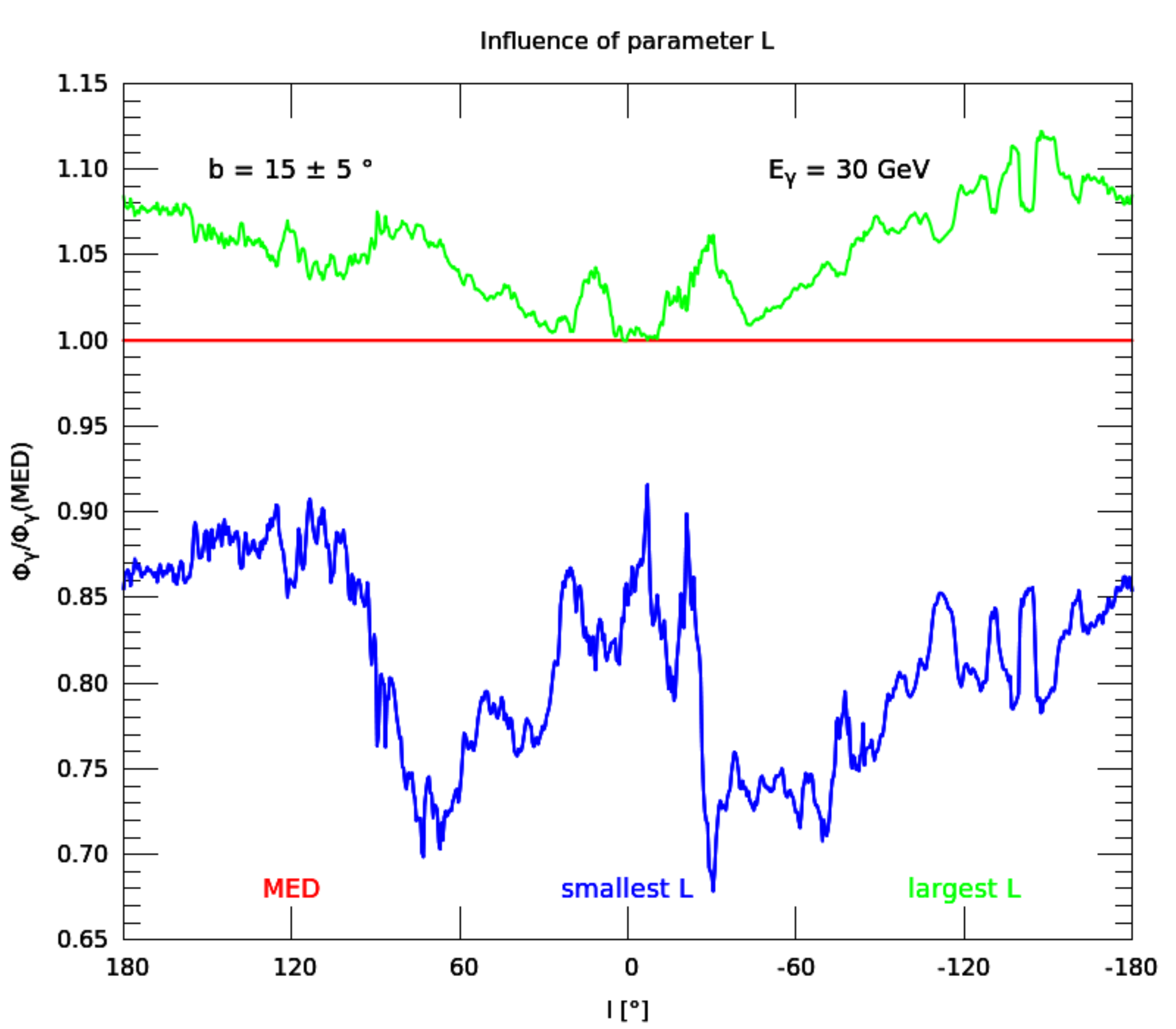}
\caption{
The half thickness $L$ of the DH was varied with respect to the MED
configuration while keeping all the other parameters fixed. The green
and blue curves correspond to the largest (15~kpc) and
smallest (1~kpc) values of $L$, respectively. The red axis stands for the reference
case (4~kpc). The variations in the $\gamma$-ray diffuse
emission are drastic. 
}
\label{fig:L_rel}
%
\vskip -0.25cm %
%
\end{figure*}

\vskip 0.1cm
We have so far varied the magnitude of Galactic CR convection or
diffusion. We have studied the interplay between these two propagation
mechanisms and showed that their impact on the $\gamma$-ray diffuse emission
is small to moderate.
We now explore how a change in the size of the DH itself would affect
the $\gamma$-ray flux. In Fig.~\ref{fig:L_rel}, the variations in the
latter are averaged over the same interval of Galactic latitude as in
Fig.~\ref{fig:b15_E30}.
When $L$ is decreased from its MED value of 4~kpc down to 1~kpc, the
thickness of the DH has considerably shrunk. Hydrogen clouds lying at
a height $|z|$ greater than 1~kpc are no longer illuminated by CR proton
and helium nuclei. The amount of gas that participates in the signal
is smaller in the MIN case than in the MED configuration although most
of the Galactic hydrogen lies within the plane of the Milky Way. As shown
by the blue
curve, the reduction in the $\gamma$-ray emission may reach 30\% in
directions where a substantial fraction of the gas lies outside the DH.
We have already witnessed that effect in the form of the dark blue spots
in Fig.~\ref{fig:minMax}. In the MIN case, the CR proton and helium
nuclei just miss the clouds located beyond the vertical boundaries of
the DH as illustrated in
Fig.~\ref{fig:minMaxwhy}.
%
%
When $L$ is increased to its MAX value of 15~kpc, the $\gamma$-ray
emission becomes stronger, as expected. However, because most of the
Galactic hydrogen lies within a distance $L_{\rm H} \leq 0.7$~kpc
from the disk,
the enhancement of the $\gamma$-ray flux is moderate and does not
exceed 8\% in the direction of the Galactic center and 12\% towards
the anti-center. The DH acts as a reservoir for Galactic hydrogen.
The larger its volume, the more gas that is illuminated by primary CR nuclei
and the larger the signal. The latter has the same variations with
$L$ as the antiproton or positron yields of the putative particles
that are supposed to make up the astronomical dark matter and to
annihilate within the Milky Way.
That is why we have featured in Table~\ref{table:prop} the typical
sets of propagation parameters borrowed from the dark matter primary
antiproton analysis of \citet{2004PhRvD..69f3501D}.

%
\begin{figure}[h!]
\centering
\includegraphics[width=0.415\textwidth]{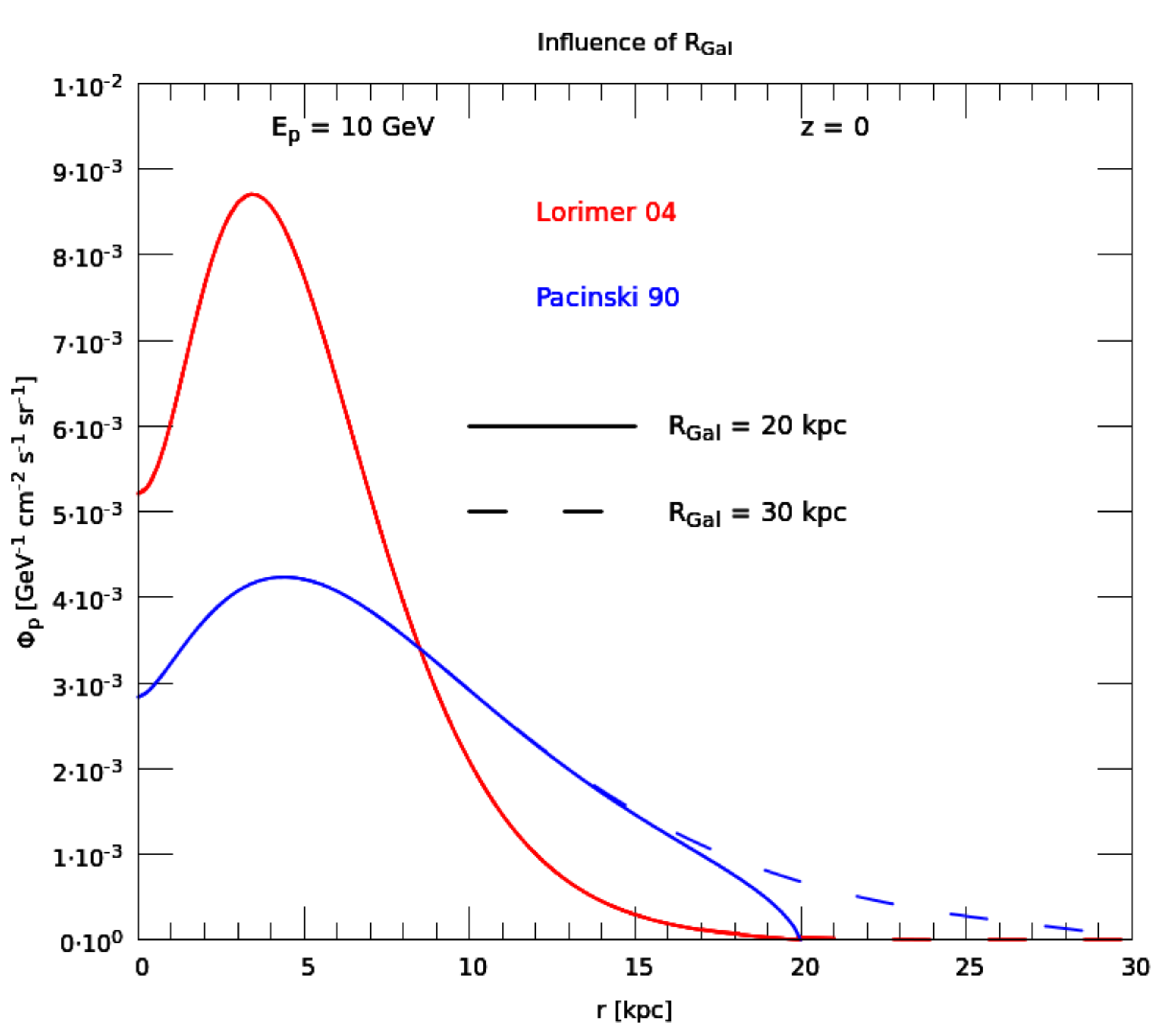}
{\hspace{0.005\textwidth}}
\includegraphics[width=0.565\textwidth]{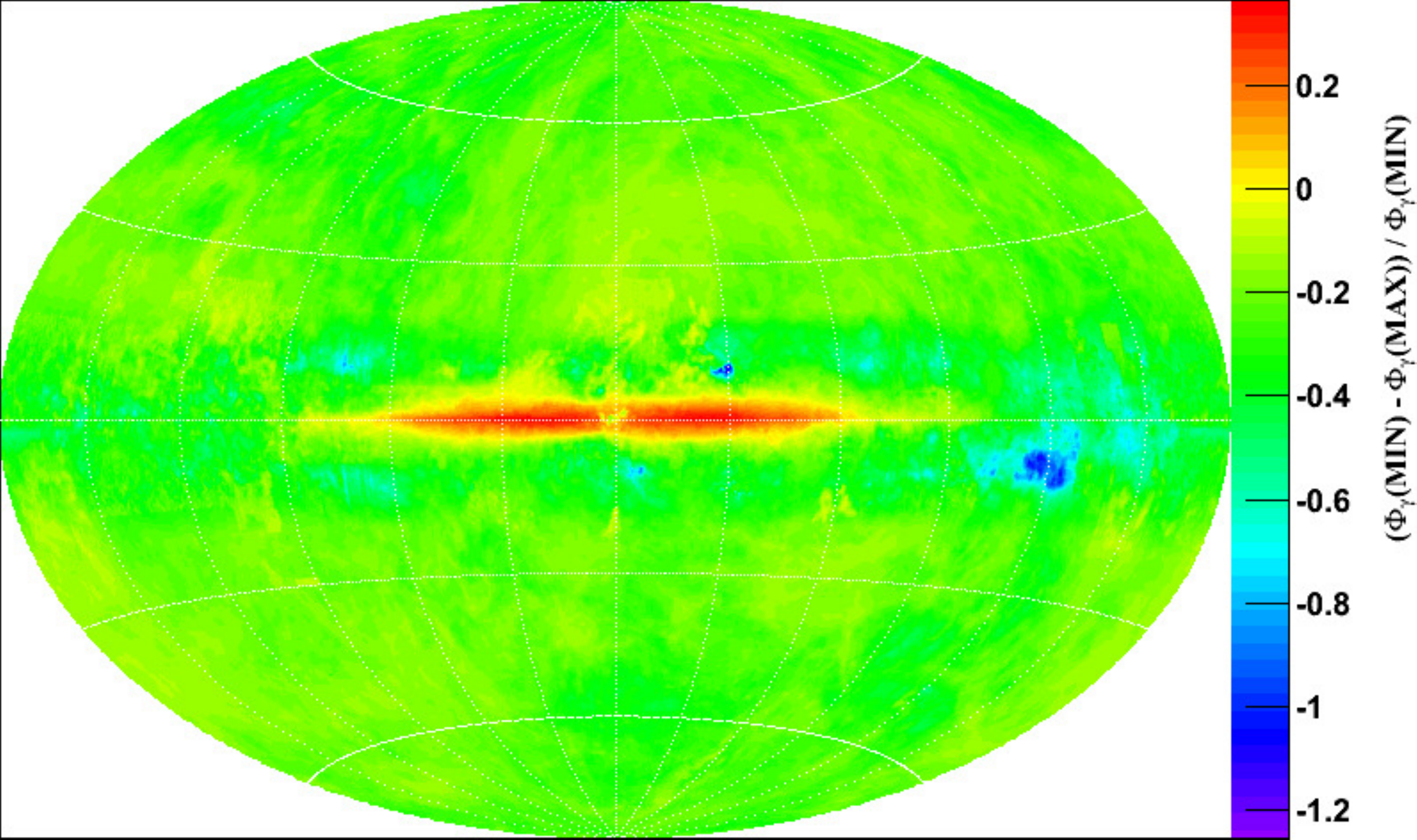}
\caption{
The influence of the radial boundary $R_{\rm Gal}$ of the DH on the Galactic
$\gamma$-ray diffuse emission:
In the left panel, the CR proton radial distribution along the Galactic disk
-- also called the CR proton gradient -- is displayed for a radial boundary
$R_{\rm Gal}$ of 20~kpc (continuous line) and 30~kpc (dashed line), for the two
source profiles L04 in red \citep{2004IAUS..218..105L} and P90 in blue
\citep{1990ApJ...348..485P}.
In the right panel, we have plotted the ratio
${\rm (map2 - map1) / map1}$ for each pixel where map1 and map2 were derived
with $R_{\rm Gal} = 20$ and 30~kpc respectively. The CR source profile is P90 here, all the
other inputs being the same as in the reference map of Fig.~\ref{fig:ref_map}.
}
\label{fig:RGAL}
%
\vskip -0.25cm %
%
\end{figure}

\vskip 0.1cm
The radial extension of the DH is the last propagation parameter whose effect
on the Galactic $\gamma$-ray diffuse emission needs to be explored.
In previous studies, we always considered $R_{\rm Gal}$ to be 20~kpc, because
CR data at the Earth are not very sensitive to this value. However, when considering
$\gamma$-rays, some variations can appear when changing the value of this parameter.
As illustrated in the right panel of Fig.~\ref{fig:RGAL}, changing $R_{\rm Gal}$
from 20 to 30~kpc induces up to a 6\%
increase in the $\gamma$-ray flux from the anti-center region. However, this result
strongly depends on the source distribution profile $\rho(r,0)$ considered.
Indeed, as shown in the left panel, if we take the source profile L04 of
\citet{2004IAUS..218..105L} (see section \ref{sec:f_SNR_radial} and
Fig.~\ref{fig:def_profil} for more details), then almost no variation comes
from the increase of $R_{\rm Gal}$. This profile gives an almost null source
density at large Galactic radii.
On the contrary, the source distribution P90 from \citet{1990ApJ...348..485P}
features a non-negligible amount of sources beyond a Galactocentric distance of
20~kpc, hence a moderate increase in the $\gamma$-ray flux from the anti-center
region.
We would like to stress that large uncertainties exist concerning the distribution
of supernova remnants at large Galactic radii. The effect of increasing
$R_{\rm Gal}$ could then be stronger than what has been derived here with the
P90 profile.

%
\section{Other sources of uncertainties}
\label{sec:other_sources_uncertainty}

In this section, we discuss first the effect of the $\gamma$-ray production
cross section and then of the injection CR proton et helium fluxes on
the effective emissivity at the Earth ${\cal E}_{\rm eff}(\odot)$ of the
ISM. We would like to disentangle the potential sources of uncertainties
that are likely to affect the $\gamma$-ray flux. That is why the discussion
of sections~\ref{sec:sigma} and \ref{sec:phi_proton_alpha_at_earth}
focuses on ${\cal E}_{\rm eff}$ and not on $\Phi_{\gamma}$.
CR Galactic propagation is involved in
section~\ref{sec:f_SNR_radial} where we analyze how various choices for
the Galactic distribution of primary CR sources influence our predictions.

%
\subsection{Gamma-ray production cross section}
\label{sec:sigma}

\begin{figure}[h!]
%
\vskip -0.25cm %
%
\centering
\includegraphics[width=0.60\textwidth]{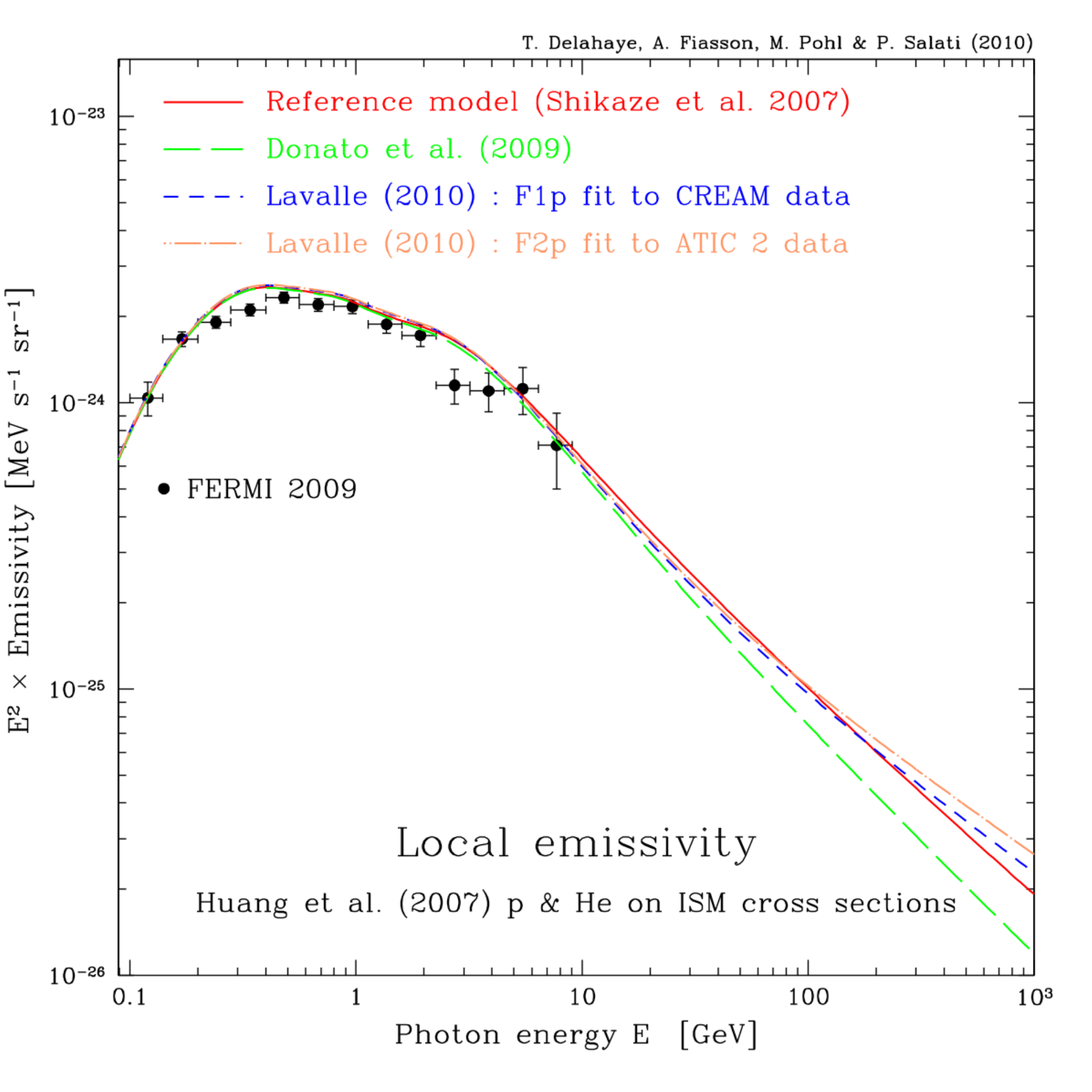}
\vskip -0.2cm
\caption{
The local effective emissivity ${\cal E}_{\rm eff}$ of the ISM per
hydrogen atom recently measured \citep{Abdo:2009ka} by the
Fermi collaboration. The black data points are plotted with various
predictions based on different $\gamma$-ray production cross sections.
The red solid curve features our reference model for which the
\citet{Huang:2006bp} cross sections are combined with the
\citet{2007APh....28..154S} proton and helium fluxes.
The magenta short-dashed curve is the emissivity for only proton-proton interactions  and is based on the~\citet{Kamae:2006bf} parameterization.
Taking then into account the nuclear weight factors of eqs.~\ref{w_NT} and \ref{w_OB} yields the blue and orange long-dashed curves. 
In all cases, the cross sections are convoluted with the CR proton and helium fluxes measured at the Earth's position by~\citet{2007APh....28..154S}.
}
\label{fig:local_emissivity}
%
\vskip -0.25cm %
%
\end{figure}

%
Most of the recent calculations of the fluxes at the Earth of
secondary CR leptons
\citep[see for instance][]{Delahaye:2008ua,2010A&A...524A..51D}
are based on the \citet{Kamae:2006bf}
parameterization of the production cross sections in proton-proton
collisions. The inclusive pion production cross section
is drawn from \citet{Blattnig:2000zf}. Older works -- in particular
the article by \citet{1998ApJ...493..694M} which has led to the GALPROP
package -- make use of the \citet{1977PhRvD..15..820B} parameterization.
The short-dashed magenta curve of Fig.~\ref{fig:local_emissivity}
has been derived with the \citet{Kamae:2006bf} parameterization and
the \citet{2007APh....28..154S} measurements of the proton flux at
the Earth.  This curve features the partial $\gamma$-ray emissivity
${\cal E}_{\rm pH}$ for which only proton-proton interactions have
been taken into account. It lies well below the other lines
for which the contributions of nuclear collisions have been
included.

%

%
The production of a $\gamma$-ray -- and of any species for that matter --
by a CR nucleus $A_{i}$ impinging on a target nucleus $A_{t}$ of the ISM
is modeled through the nuclear weight factor $w$ defined as
\beq
{\displaystyle \frac{d{\sigma}}{dE}}
(A_{i}[T] + A_{t} \to \gamma[E]) \, = \,
w(A_{i} , A_{t} , T , E) \, \times \,
{\displaystyle \frac{d{\sigma}}{dE}}
(p[T] + H \to \gamma[E]) \;\; .
\label{def:nuclear_weight}
\eeq
The weight $w$ gauges the nuclear effects involved in the
$\gamma$-ray production during a nucleus-nucleus interaction and
scales them with respect to the equivalent proton-proton collision.
It depends on the atomic numbers $A_{i}$ and $A_{t }$ of the projectile
and target nuclei. Its variations with the energy $E$ of the
photon and the kinetic energy per nucleon $T$ of the CR nucleus
are in general disregarded. A nucleus-nucleus collision can be
naively understood as a collection of nucleon-nucleon interactions
where protons and neutrons have identical behaviors.
Relation~(\ref{def:nuclear_weight}) is based on this scheme and
$w$ accounts for the number of two body reactions occurring
during a nuclear collision. Various prescriptions have been proposed
in the literature such as the \citet{1976ApJ...206..312O} formula
\beq
w_{\rm OB} \, = \, \left(
A_{i}^{3/8} \, + \, A_{t}^{3/8} \, - 1 \right)^{2} \;\; ,
\label{w_OB}
\eeq
or the more recent \citet{Norbury:2006hp} scaling factor
\beq
w_{\rm NT} \, = \, \left( A_{i} \, A_{t} \right)^{0.8} \;\; .
\label{w_NT}
\eeq
Both prescriptions have been combined with the \citet{Kamae:2006bf}
proton-proton cross section and the \citet{2007APh....28..154S}
proton and helium fluxes to yield the long-dashed dotted orange
(KOB) and the long-dashed blue (KNT) curves, which are close to. The \citet{1976ApJ...206..312O} nuclear
weight leads to a result that is slightly smaller than the
\citet{Norbury:2006hp} prediction. The KOB and KNT models actually differ
by less than 3.7\% for a photon energy $E$ in the range
extending from 100~MeV to 1~TeV.
The red curve corresponds to our reference model. The emissivity
has been obtained by combining the \citet{Huang:2006bp} cross sections
with the \citet{2007APh....28..154S} proton and helium fluxes. The
result is significantly higher than for the KNT or KOB models,
with a difference of about 33\% at 1~GeV, which reaches a maximum
of 54\% at 4.5~GeV and decreases down to 20 to 30 \% above 100~GeV.
Because the \citet{Huang:2006bp} parameterization is based on the
\citet{Kamae:2006bf} proton-proton cross section at low proton
energy, our reference model is quite close to the KNT or KOB
predictions at low photon energy. The difference is only 13\% at
100~MeV. At high energy, the DPMJET-III code incorporates the
direct production of photons, a process which is not included
in the \citet{Kamae:2006bf} parameterization. Nuclear reactions
are also modeled more accurately.

\vskip 0.1cm
In the literature, the nuclear effects are often described by the nuclear enhancement factor $\epsilon_{M}$, which is defined as the ratio
\beq
\epsilon_{M} \, = \, {\displaystyle
{{\cal E}_{\rm eff} (\mathbf{x},E)}/
{{\cal E}_{\rm pH}  (\mathbf{x},E)}} \;\; .
\eeq
This factor depends a priori on the photon energy $E$ and on 
position $\mathbf{x}$ within the CR diffusive halo. Even though
the various nuclear weight factors $w$ may only depend on the atomic
numbers of the projectile and target nuclei, the CR helium and
proton fluxes do not undergo the same variations as the kinetic
energy per nucleon $T$ and the position $\mathbf{x}$ are changed.
We calculated $\epsilon_{M}$ in the KNT model and mapped its
variations inside the CR diffusive halo for different photon
energies. We find that $\epsilon_{M}$ is basically constant.
At the sun position, it decreases from 1.544 at 1~GeV down to
1.526 at 100~GeV. Throughout the CR diffusive halo, $\epsilon_{M}$
varies by a factor of 2.4\% at 1~GeV. The amplitude of the spatial
variations decreases to only 0.15\% at 100~GeV. We find similar
results for the KOB model, with a solar value of
$\epsilon_{M}$ decreasing from 1.491 at 1~GeV down to 1.475
at 100~GeV.
A good estimate of the nuclear enhancement factor is therefore given
by its local value. Assuming that relations~(\ref{w_OB}) or (\ref{w_NT})
hold leads to the approximate expression
\beq
\epsilon_{M} \, = \, {\displaystyle \sum_{\rm A}} \;
{\displaystyle \frac{X_{\rm A}}{X_{\rm H}}} \; \left\{
w(1,{\rm A}) \; + \; w(4,{\rm A}) \,
{\displaystyle \frac{\Phi_{\alpha}(\odot)}{\Phi_{p}(\odot)}}
\right\} \;\; ,
\label{simple_epsilon_M}
\eeq
where the helium to proton CR flux ratio at the Earth has been
taken as constant.
This is not quite correct actually. Because the
\citet{2007APh....28..154S} spectral indices for CR proton and
helium nuclei are equal to 2.76 and 2.78 respectively, we get
a small decrease of $\epsilon_{M}$ with energy.
The element abundances in the ISM also matter. If we adopt the solar
value of 0.0975 for the ${X_{\rm He}}/{X_{\rm H}}$ ratio instead of
the interstellar canonical value of 0.111, the KNT nuclear enhancement
factor decreases by 3\% and is now very close to the KOB value.

\vskip 0.1cm
The predictions of the KNT and KOB models at 10~GeV -- respectively
1.535 and 1.483 -- are in very good agreement with the value of
1.52 found by \citet{1992ApJ...394..174G}.
Since that publication, the modeling of nuclear interactions has
been considerably improved thanks in particular to the DPMJET
event generator.
But because the \citet{Huang:2006bp} parameterization is given
for proton and helium nuclei interactions with the ISM taken as
a whole, the nuclear enhancement factor $\epsilon_{M}$ cannot be
extracted. We had no access to the \citet{Huang:2006bp} proton-proton
cross section and could not compute ${\cal E}_{\rm pH}$. To get
a feeling for how large $\epsilon_{M}$ can be in that case, we used
the nuclear weight factors $w$ that \citet{Mori:2009te} calculated
with the DPMJET-III Monte Carlo code and we combined them with the
\citet{Kamae:2006bf} proton-proton cross section and the same
ISM composition as in \citet{Huang:2006bp}. At 10~GeV, we derive
a value of 1.69, well above the KNT or KOB predictions.
\citet{Mori:2009te} finds a nuclear enhancement factor
as large as 1.84 at 10~GeV, increasing up to 2.00 at 1~TeV. In his
analysis, nuclei up to Fe are included in both the ISM and the CR
projectiles. The parameterization of the various CR fluxes is also
borrowed from \citet{Honda:2004yz} whereas our estimate of
$\epsilon_{M}$ relies on the \citet{2007APh....28..154S} measurements.
The CR helium-to-proton flux ratio increases like $T^{0.1}$ in the
former case, whereas it decreases slightly in our case.
We never made use of this nuclear enhancement factor $\epsilon_{M}$ because our calculations are based directly on $\gamma$-ray production cross-sections, nuclear weight factors $w$, and the CR proton and helium fluxes measured at the Earth.

%
\subsection{Cosmic ray proton and helium fluxes at the Earth}
\label{sec:phi_proton_alpha_at_earth}

\begin{figure}[h!]
%
\vskip -0.25cm %
%
\centering
\includegraphics[width=0.60\textwidth]{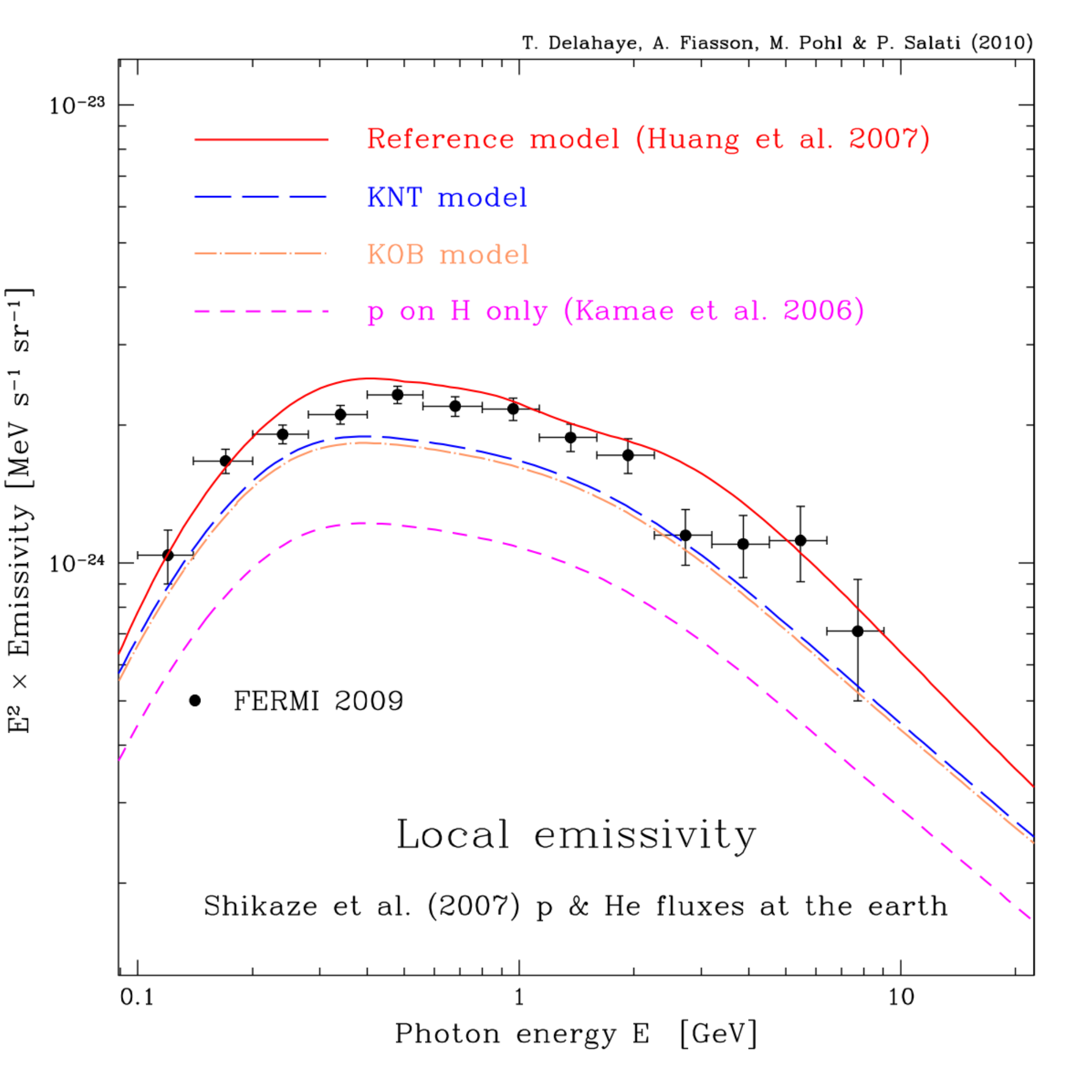}
\vskip -0.2cm
\caption{
The measurements by \citet{Abdo:2009ka} of the local effective
emissivity ${\cal E}_{\rm eff}$ of the ISM per hydrogen atom
(black data points) are compared to various predictions for
which the \citet{Huang:2006bp} cross sections have been used.
A selection of different parameterizations for the CR proton
and helium fluxes at the Earth is featured.
The curves diverge from each other above 10~GeV and the spread
among them reaches a factor of 2 at 1~TeV.
}
\label{fig:emis_p_alpha_fluxes}
%
\vskip -0.25cm %
%
\end{figure}

The fluxes at the Earth of CR protons and helium nuclei actually do matter
in calculating ${\cal E}_{\rm eff}$. In
Fig.~\ref{fig:emis_p_alpha_fluxes}, the \citet{Huang:2006bp}
cross sections have been combined with a selection of different
parameterizations to yield the ISM effective emissivity at the
solar circle.
The red solid curve stands for our reference model based on the
\citet{2007APh....28..154S} proton and helium fluxes. It extends up
to 1~TeV the corresponding curve of Fig.~\ref{fig:local_emissivity}.
Because hadronic interactions are basically scale invariant, the
emissivity decreases with roughly the same spectral index as the
progenitor CR proton and helium fluxes. The latter have been
parameterized by \citet{2007APh....28..154S} with the simple
power law
\beq
\Phi \, = \, A \; \beta^{p_{1}} \; {\cal R}^{- {p_{2}}} \;\;
{\rm cm}^{-2} \, {\rm s}^{-1} \, {\rm sr}^{-1} \,
({\rm GeV/n})^{-1} \;\; ,
\label{eq:flux_fit}
\eeq
where ${\cal R}$ denotes the rigidity of the nucleus expressed in units
of GV. The
\citet{2007APh....28..154S} measurements yield the
parameterizations
$(A , p_{1} , p_{2}) = (1.94 , 0.7 , 2.76)$ for H and
%
%
$(0.71 , 0.5 , 2.78)$ for He.
%
%
Between 3~GeV and 1~TeV, our reference model leads to an emissivity
approximated well by
\beq
{\cal E}_{\rm eff}({\odot},E) \simeq 3.55 \times 10^{-27} \;
{\rm GeV^{-1} \, s^{-1} \, sr^{-1}} \;
\left(
{\displaystyle \frac{1 \, {\rm GeV}}{E}}
\right)^{2.76} \;\; .
\eeq

%

\vskip 0.1cm
A crucial element in the calculation of the Galactic diffuse
$\gamma$-ray emission is the interstellar (IS) fluxes of protons and helium
nuclei. These fluxes are actually measured at the top of the terrestrial
atmosphere (TOA) and have been considerably altered during their journey
through the heliosphere.
As pointed out by~\citet{Donato:2008jk}, determination of the
spectral indices of the IS fluxes from their TOA values is
potentially spoilt by solar modulation. The latter tends to harden the
spectra even beyond 10~GeV/n so that only the highest energy points should
be used to parameterize the IS fluxes. Unfortunately, these points
are most subject to statistical uncertainties as the CR fluxes decrease
at high energy where very few events are collected. CR measurements with large
detectors will ensure in the future substantial statistics and will help
determine the IS high-energy spectra with improved accuracy.
To gauge this effect, we used the \citet{Donato:2008jk}
parameterizations, which are borrowed from \citet{2007APh....28..154S}
below 20~GeV/n and are otherwise given by
$(A , p_{1} , p_{2}) = (2.4132 , 0 , 2.839)$ for H and
%
%
$(0.8866 , 0 , 2.85)$ for He.
%
%
This yields the curve that departs from the red
solid line above 10~GeV and is a factor 0.62 below it at 1~TeV.

\vskip 0.1cm
The balloon experiments ATIC2 \citep{atic2:2009} and CREAM \citep{Ahn:2010gv}
have recently reported a significant hardening of the proton and helium spectra
above 200~GeV/n. Although both experiments agree on the helium spectrum,
ATIC2 finds a larger proton excess than CREAM. These new fluxes have been
parameterized by \citet{Lavalle:2010sf} and lead to those two other curves. The former is based on the
\citet{Lavalle:2010sf} F1p fit to the proton data measured by CREAM, while
the latter makes use of the F2p fit to the ATIC2 proton flux. Once again,
the various emissivities show similar behaviors below 10~GeV. Up
to 100~GeV, the F1p and F2p predictions are below the reference model
whose parameterization of the proton and helium fluxes does not reproduce
the dip observed in the ATIC2 and CREAM data well. Above 100 to 200~GeV,
the F1p and F2p emissivities overreach the reference model by a factor
respectively equal to 1.19 and 1.37 at 1~TeV.
This is comparable to the increase of about 30\% found by \citet{2011PhRvD..83b3014D} for photons at energies close to 300~GeV.

%
\subsection{Distribution of the primary cosmic ray sources in the disk}
\label{sec:f_SNR_radial}

Though convincing proof remains to be given, it is widely accepted that
the main source of CR proton and $\alpha$ particles in the GeV-TeV
energy range are the supernova remnants. Indeed there are more and more
observational hints pointing towards supernova remnants -- see for
instance~\citet{2009AIPC.1112...54F}. From the theoretical point of view,
these objects are the perfect location for Fermi acceleration. However,
supernova remnants are very difficult to detect because as they get older,
they expand hence become dimmer and dimmer, so that even nearby ones
are difficult to see because they sometimes become so big in the sky that they
exceed the angular acceptance of the radio telescopes. The most up-to-date
catalog \citep{2009BASI...37...45G} contains only 274 objects and the precise
location of most of them is unknown. However, as two thirds of the
supernov\ae~are expected to have undergone core collapse, one can use
pulsars as tracers
of the supernova remnants distribution. The ATNF
catalog\footnotemark~\citep{2005AJ....129.1993M} lists more than 1800 pulsars.
Nevertheless, too naive use of the statistics would lead to errors, since
it is well known that data do not reflect reality faithfully because of
detection biases~\citep[{e.g.}][]{2004IAUS..218..105L}.
\footnotetext{\url{http://www.atnf.csiro.au/research/pulsar/psrcat}}

%
%
\vskip -0.25cm %
%
\begin{figure}[h!]
\centering
\includegraphics[width=0.6\textwidth]{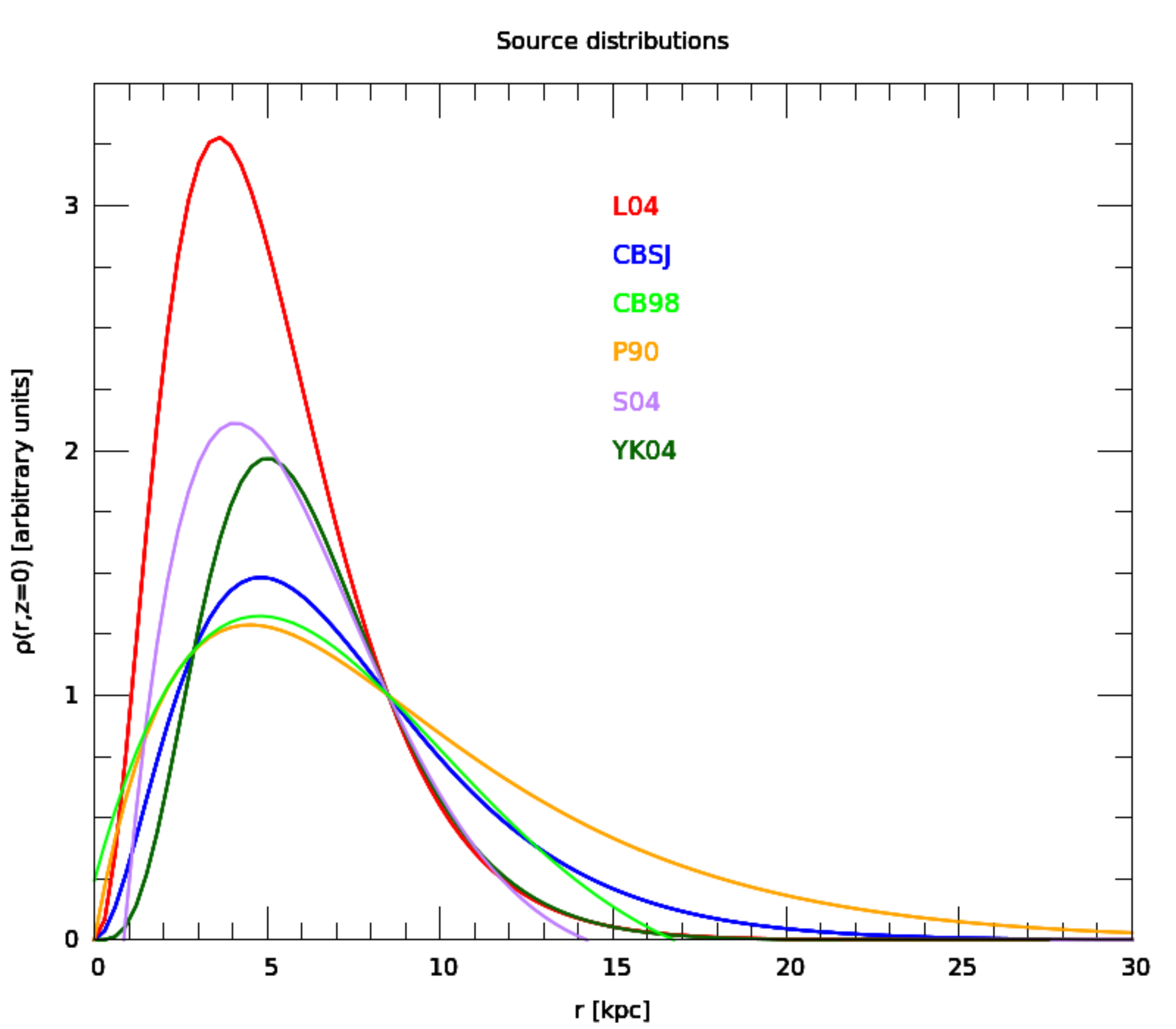}
\caption{
Various cosmic ray source (supernova remnants) distributions are available
in the literature. They are plotted here as a function of the Galactocentric
distance $r$. The references of the curves are
\citet{2004IAUS..218..105L} in red (L04),
\citet{1998ApJ...504..761C} in blue (CBSJ),
\citet{1998ApJ...504..761C} in light green (CB98),
\citet{1990ApJ...348..485P} in orange (P90),
\citet{2003ICRC....5.2639S} in purple (S04) and
\citet{2004A&A...422..545Y} in dark green (YK04).
For clarity, all these distributions have been normalized here
with respect to their solar circle values. However, this is not what is
used in our calculations since the normalization of the CR production rates
are set automatically by the \textit{retropropagation} method explained
in section~\ref{sec:CR_model}.
}
\label{fig:def_profil}
%
\vskip -0.25cm %
%
\end{figure}

\vskip 0.1cm
There are few distribution models in the literature that we can compare.
Most of them exhibit radial dependencies in the form proposed
by~\citet{1977ApJ...217..843S} with
\begin{equation}
\rho(r,z) \, = \, \rho_0 \times r^{\displaystyle a} \times
\exp \left\{ - \frac{r}{r_0} \right\} \;\; ,
\label{eq:rho}
\end{equation}
where $\rho_0$ is the normalization\footnote{This normalization is not very
important for us as our \textit{retropropagation} method automatically sets
the magnitude of the CR Galactic production rates $Q_{\rm tot}$ in order to
ensure the correct CR fluxes at the Earth.}
and $r$ the distance from the Galactic center. Having made use of the thin
disk approximation to describe the Galactic plane, we have not considered
any  vertical variation of these profiles.
Different sets of values can be found in the literature for the pair of
parameters $(a , r_0)$.
\citet{2004IAUS..218..105L}, hereafter L04,  found   $(2.35 \, , \, 1.528 \, {\rm kpc})$ and
\citet{2004A&A...422..545Y}, hereafter YK04, derived $(4    \, , \, 1.25  \, {\rm kpc})$.
\citet{1998ApJ...504..761C}, hereafter CBSJ, got     $(2.0  \, ,  \, 3.53 \, {\rm kpc})$ while
\citet{1990ApJ...348..485P}, hereafter P90, determined
$(1 \, , \, 4.5 \, {\rm kpc})$.
Finally, at variance with the parameterization sketched above, we mention the
distribution proposed by~\citet{1998ApJ...504..761C} and denoted hereafter CB98.
Though obtained from a fit performed on a poor statistical ensemble of only 36
supernova remnants, that profile has the particularity of giving a non zero density
at the Galactic center since
\begin{equation}
\rho(r,0) \, = \, \rho_0 \times
\sin \left( \pi \frac{r}{r_s} + \theta \right) \times 
\exp \left\{ - \frac{r}{r_0} \right\} \;\; .
\end{equation}
The authors find that
$r_0 = 7.7  \pm 4.7$~kpc,
$r_s = 17.2 \pm 1.9$~kpc and
$\theta = 0.08 \pm 0.33$. This relation is only valid as long as
$r \leq r_s \times (1 - {{\theta}/{\pi}})$ ({\ie} in the inner
16.8~kpc) and is zero beyond.
All these distributions are displayed in Fig.~\ref{fig:def_profil}.
\citet{2006ApJ...639L..25B} have recently reported the detection of
35 new supernova remnants in the inner Galaxy and suggest that earlier radial
distribution estimations should be revised.

%
\begin{figure}[h!]
\centering
\includegraphics[width=0.45\textwidth]{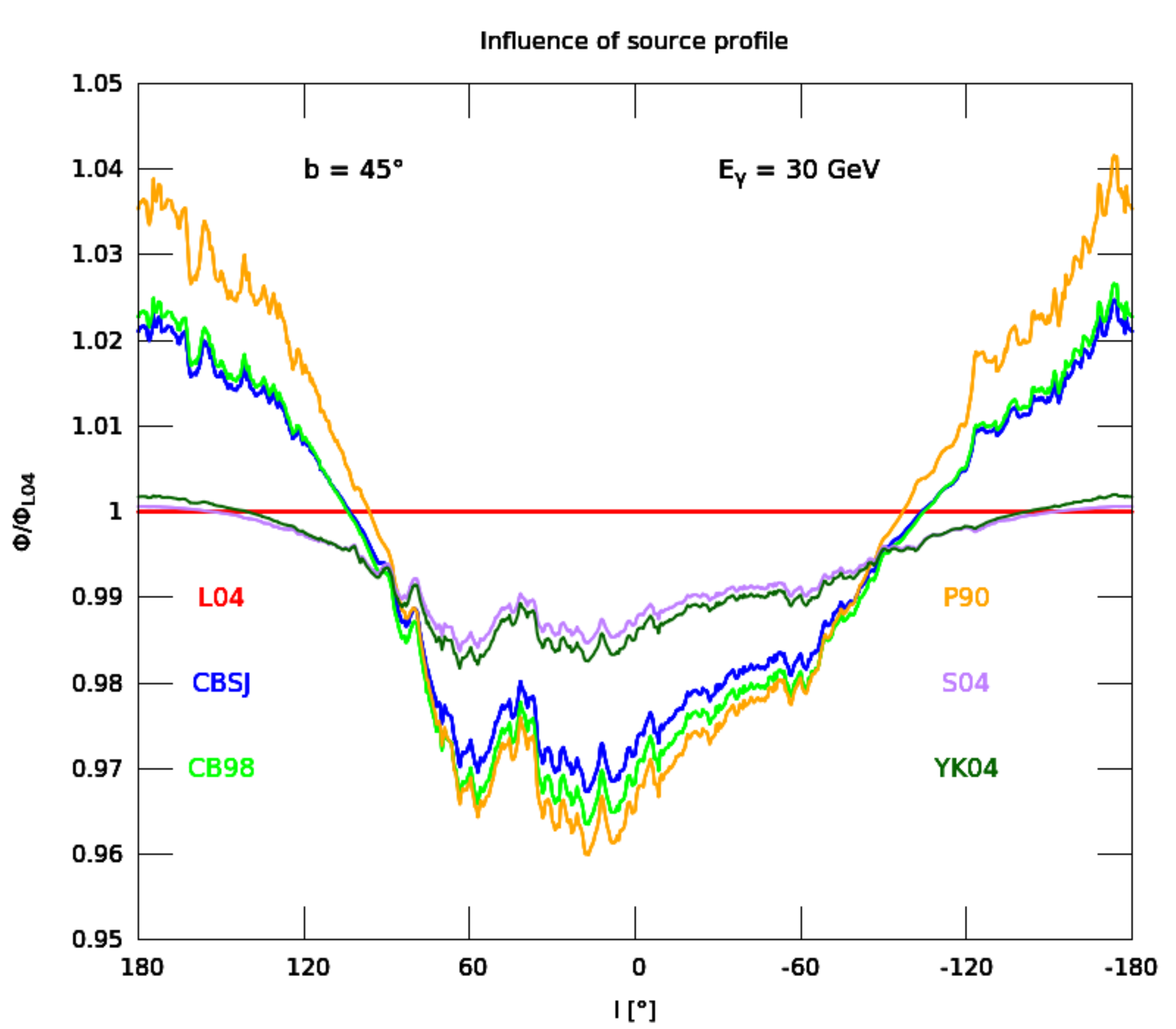}
\includegraphics[width=0.45\textwidth]{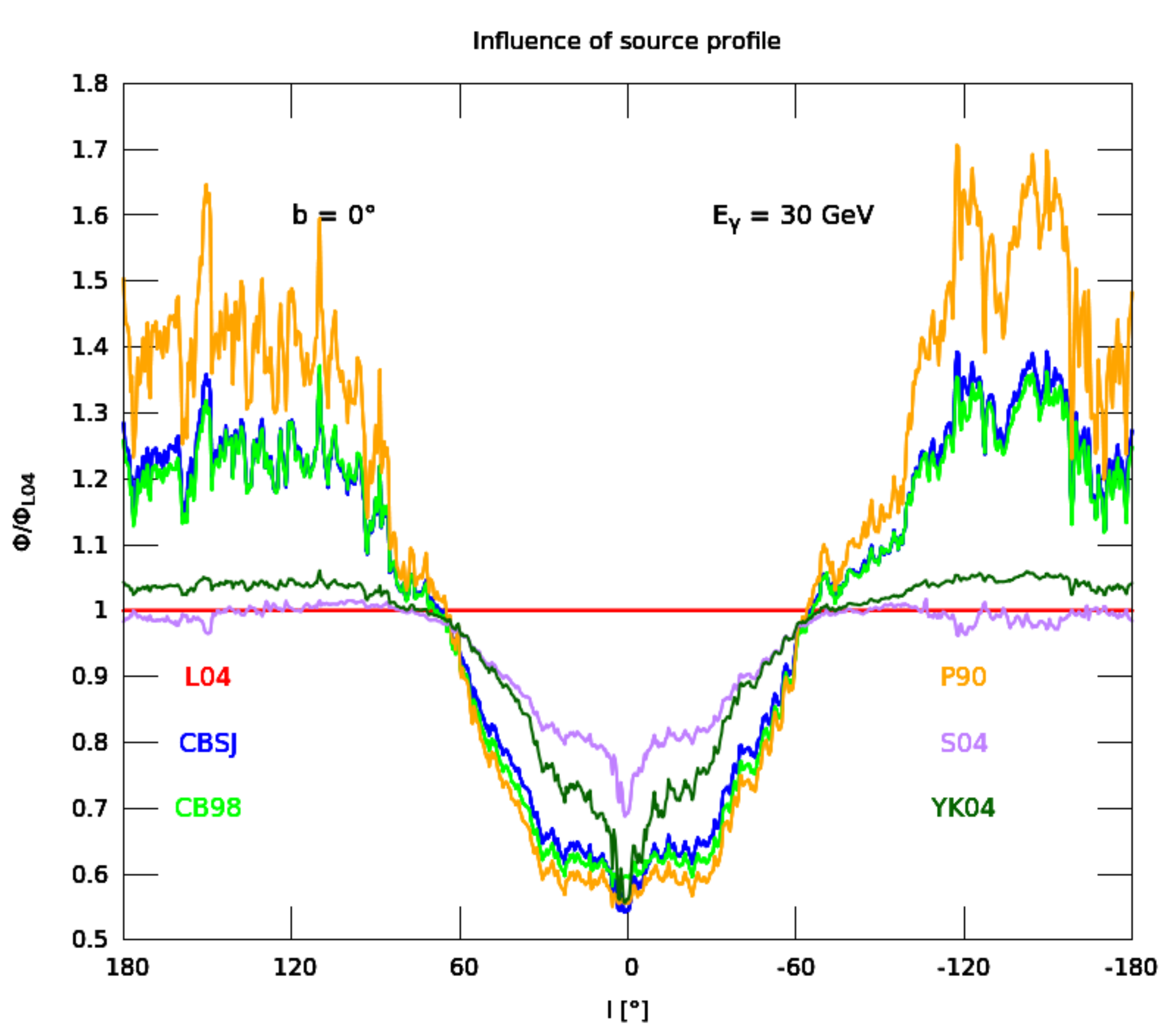}
\caption{
Variations in the $\gamma$-ray flux as a function of the Galactic
longitude $l$ relative to the reference model L04 of
Fig.~\ref{fig:ref_map}. All the source distributions described in
section~\ref{sec:f_SNR_radial} are displayed. The left panel
corrresponds to the Galactic disk with a latitude $b$ of $0^{\circ}$,
whereas the right panel features the case of the intermediate
latitude $b = 45^{\circ}$.
}
\label{fig:profil}
%
\vskip -0.25cm %
%
\end{figure}

\vskip 0.1cm
As is clear from Fig.~\ref{fig:def_profil}, there are large variations
from one model to another. The position and the magnitude of the peak
of the supernova remnant distribution varies from one author to another.
Even more striking, far away from the Galactic center, for a distance $r$
greater than 12~kpc, the distributions are very different. For instance,
the CB98 profile has no source beyond 17~kpc, whereas in the P90 case,
the source distribution extends far away beyond 20~kpc. As already
mentioned in section~\ref{sec:cr_propagation_error} and evoked in
the right panel of Fig.~\ref{fig:RGAL}, the differences in the radial
profiles of the primary CR sources from one author to another imply
large variations in the expected $\gamma$-ray flux in the anti-center
direction.
As illustrated in Fig.~\ref{fig:profil}, the uncertainties on
the $\gamma$-ray diffuse emission due to the CR source distribution can
be as large as 70\% towards the Galactic anti-center and reach up to 50\%
in the direction of the Galactic center. However, because diffusion is quite
efficient, the memory of the source distribution in the cosmic ray gradient
is erased quite quickly away from the Galactic plane. Indeed, at the
latitude $b$ of $45^{\circ}$ presented in the right panel of
Fig.~\ref{fig:profil}, the uncertainties only decrease to 4\%.

%
\section{The distribution of hydrogen in the Galaxy}
\label{sec:galactic_H}

The hydrogen density $n_{\rm H} (\mathbf{x})$ is a key ingredient in
calculating the $\gamma$-ray diffuse emission flux $\Phi_{\gamma}$.
Each hydrogen atom nucleus or proton of the ISM that lies inside the DH
is embedded inside radiations of cosmic protons and $\alpha$ particles,
and shines in the $\gamma$-ray band. Determining the Galactic
distribution of hydrogen  with accuracy is therefore paramount. This
distribution falls into three distinct parts of unequal importance.
Atomic neutral hydrogen HI is directly traced through its 21-cm line
emission. It has been mapped all over the sky and an up-to-date survey
is provided by the \citet{2005A&A...440..775K} catalog.
Molecular hydrogen is the dominant component inside the Galactic disk.
It acts as a coolant that triggers gas collapse and stellar formation.
It cannot be detected directly but its presence is traced by molecules,
such as CO, which are associated to stellar activity. The most up-to-date
map is the composite CO survey by \citet{2001ApJ...547..792D}.
Finally, ionized hydrogen HII has been included in our calculations,
although it is an inessential component.

%
\subsection{The hydrogen 3D distribution of this analysis}
\label{sec:pohl_3D}

%

We used here the results of \citet{2008ApJ...677..283P}, who
have kinematically deconvolved the composite \citet{2001ApJ...547..792D}
CO survey with a gas-flow model derived from smoothed particle hydrodynamics
(SPH) simulations in gravitational potentials based on the NIR luminosity
distribution of the bulge and disk \citep{2003MNRAS.340..949B}. Besides
providing a more accurate picture of cloud orbits in the inner Galaxy,
a fundamental advantage of this model is that it provides kinematic
resolution towards the Galactic center, in contrast to standard
deconvolution techniques based on purely circular rotation.
The same technique was applied to deconvolve the LAB HI survey
\citep{2005A&A...440..775K} with appropriate modifications reflecting
the larger single-cloud linewidth and the galactic warp and flaring
in the outer Galaxy \citep{2009ARA&A..47...27K}. The kinematic
deconvolution is applied only in the Galactic plane, whereas both the
CO and HI data are distance-allocated according to scale height at
$\vert b \vert >  5^{\circ}$ and
$\vert b \vert > 10^{\circ}$, respectively. Any kinematic
deconvolution suffers from systematic errors, and artifacts may appear
in the final maps. For example, in some regions line signal is observed
at atypical velocity for some reason, which the code would allocate in
distance according to the global flow model. It is rare that such
features dominate the gas maps, but we cannot exclude that some peaks
in the difference map (see Fig.~2) are enhanced by these systematic
effects.

\vskip 0.1cm
No systematic correction is made for HI absorption or self-absorption.
The former is potentially important in areas of high 21-cm continuum
intensity, and we have indeed interpolated  (in $l$, $b$) the spectra
towards the Galactic center or Cas A to correct negative spikes, but
a general and systematic correction would very much depend on the
assumed hydrogen spin temperature \citep{2003ApJ...585..801D}.
HI self-absorption arises from cold clouds located in front of warmer
hydrogen and is generally a small-scale feature that may be expected
to be washed out when considering large-scale features
\citep{2000ApJ...540..851G}.

%
\subsection{The GALPROP gas maps}
\label{sec:galprop}

Based on the surveys of \citet{2005A&A...440..775K} for HI
and \citet{2001ApJ...547..792D} for CO, and making use of various
models for the rotational curves of the Galaxy, which are detailed
in the appendix of \citet{2002ApJ...565..280M}, Seth W. Digel
built a gas distribution map for the GALPROP
package~\citep{1998ApJ...509..212S}, which is available
online. This model is based on a cylindrical symmetry
(the Galaxy is described as 8 annuli) and does not take
the spiral arm structure of the Milky Way into account.

%
\begin{figure}[h!]
\centering
\includegraphics[width=\textwidth]{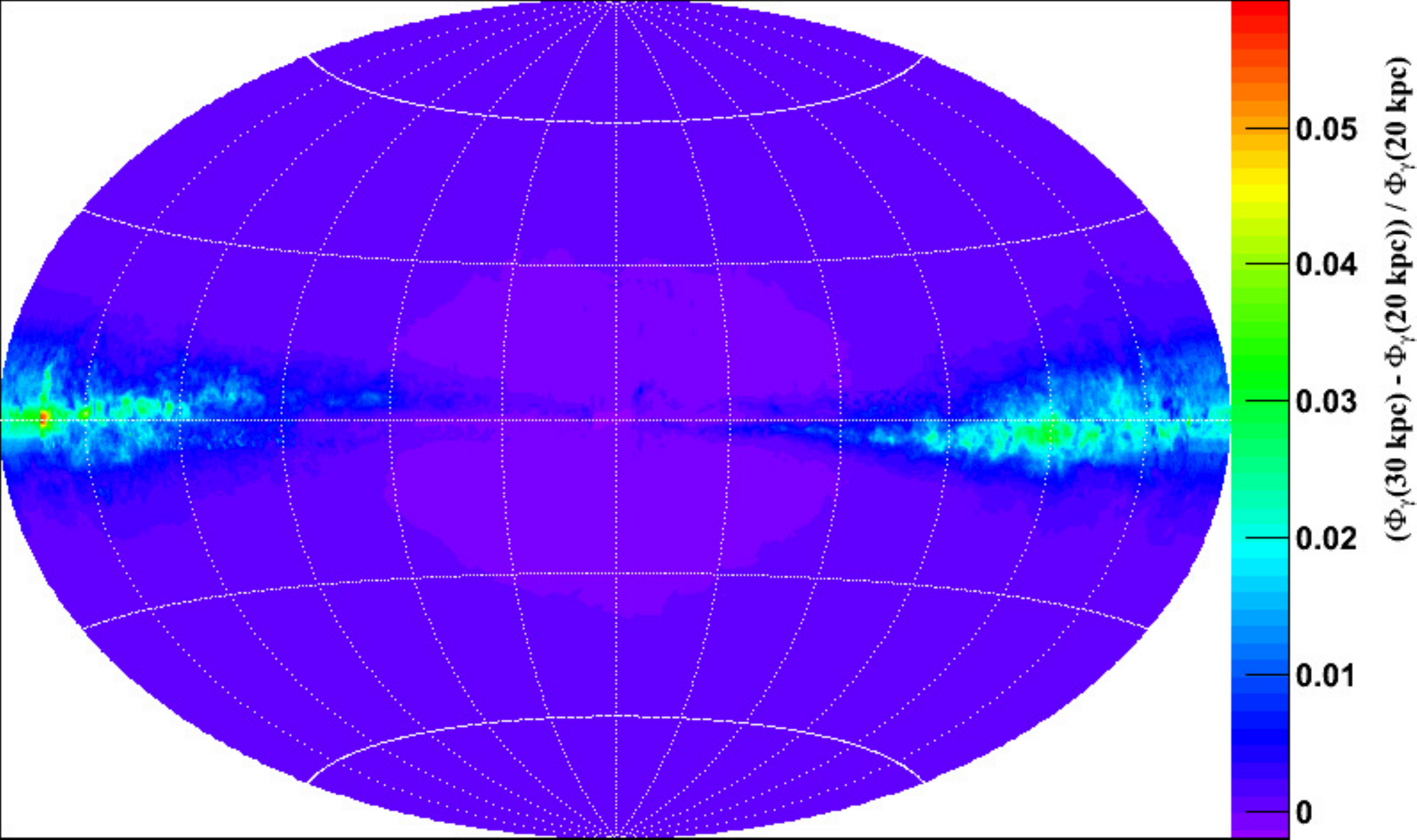}
\caption{
Relative variations in the Galactic $\gamma$-ray diffuse emission with
respect to the reference case of Fig.~\ref{fig:ref_map}. We have plotted
here the ratio
${\rm {(map1 - map2)}/{map1}}$
where map1 is our reference map whereas map2 was obtained from the
hydrogen distribution available in the GALPROP package with all the other
inputs the same as for map1. In particular, the photon energy is
30~GeV.
}
\label{fig:galprop}
%
\vskip -0.25cm %
%
\end{figure}

\vskip 0.1cm
Fig.~\ref{fig:galprop} displays the relative variations in the
$\gamma$-ray flux when using the GALPROP gas maps instead of the
\citet{2008ApJ...677..283P} distributions.
In order to disentangle the effect of the actual
3D gas distribution from the influence of the $X_{\rm CO}$ factor,
which will be discussed in the next section, we took a constant
value for the latter (as in the original GALPROP work).
Also, we have not compared the output from the GALPROP
routine to our $\gamma$-ray flux here. We have merely incorporated the GALPROP
hydrogen maps in our calculations to assess the influence of the sole
Galactic gas distribution. We therefore prepared a $\gamma$-ray
sky map in exactly the same conditions as for the reference map of
Fig.~\ref{fig:ref_map} except for the hydrogen density, which was
borrowed from the GALPROP package.
The relative differences between the GALPROP-inspired map (map2) and
our reference case (map1) have been plotted in Fig.~\ref{fig:galprop},
which clearly points toward strong contrasts.
The blue spots below the Galactic plane
are caused by different treatments of the Magellanic
Clouds and Messier 31.
Some pixels have values higher than unity because the map computed with
the GALPROP gas distribution unexpectedly leads to negative results
in some directions. The other structures that appear are probably caused by
the annular structure of the GALPROP gas map.
Finally, on top of the CO maps, an inspection of Fig.~\ref{fig:galprop}
at high Galactic latitude indicates that the HI maps are also quite
different. Taking the spiral arm structure of the
Milky Way into account is extremely important and has a strong effect on the predictions
of the Galactic $\gamma$-ray diffuse emission.

%
\subsection{The $X_{\rm CO}$ factor}
\label{sec:X_CO}

\begin{figure}[h!]
\centering
\includegraphics[width=.45\textwidth]{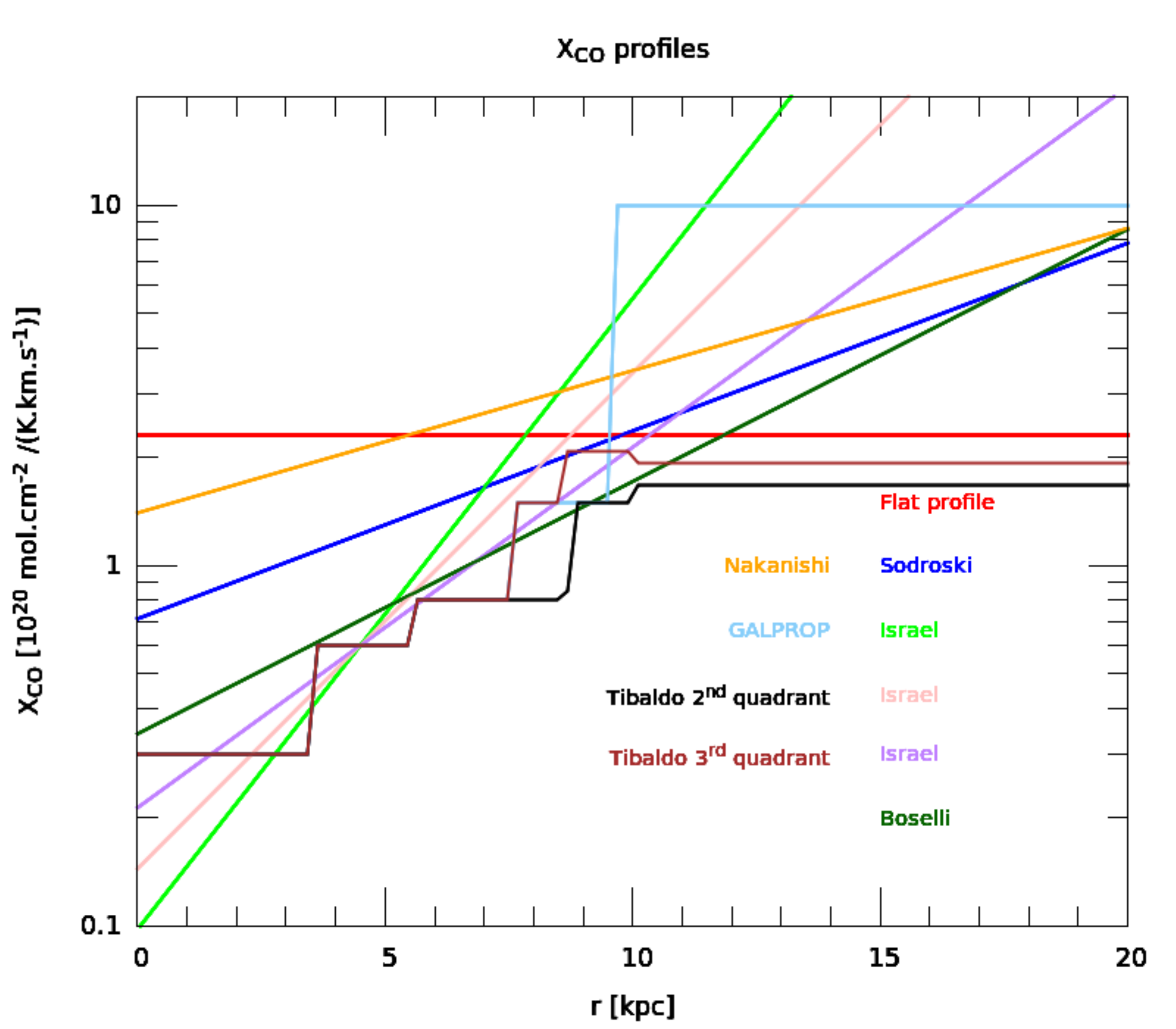}
\includegraphics[width=.45\textwidth]{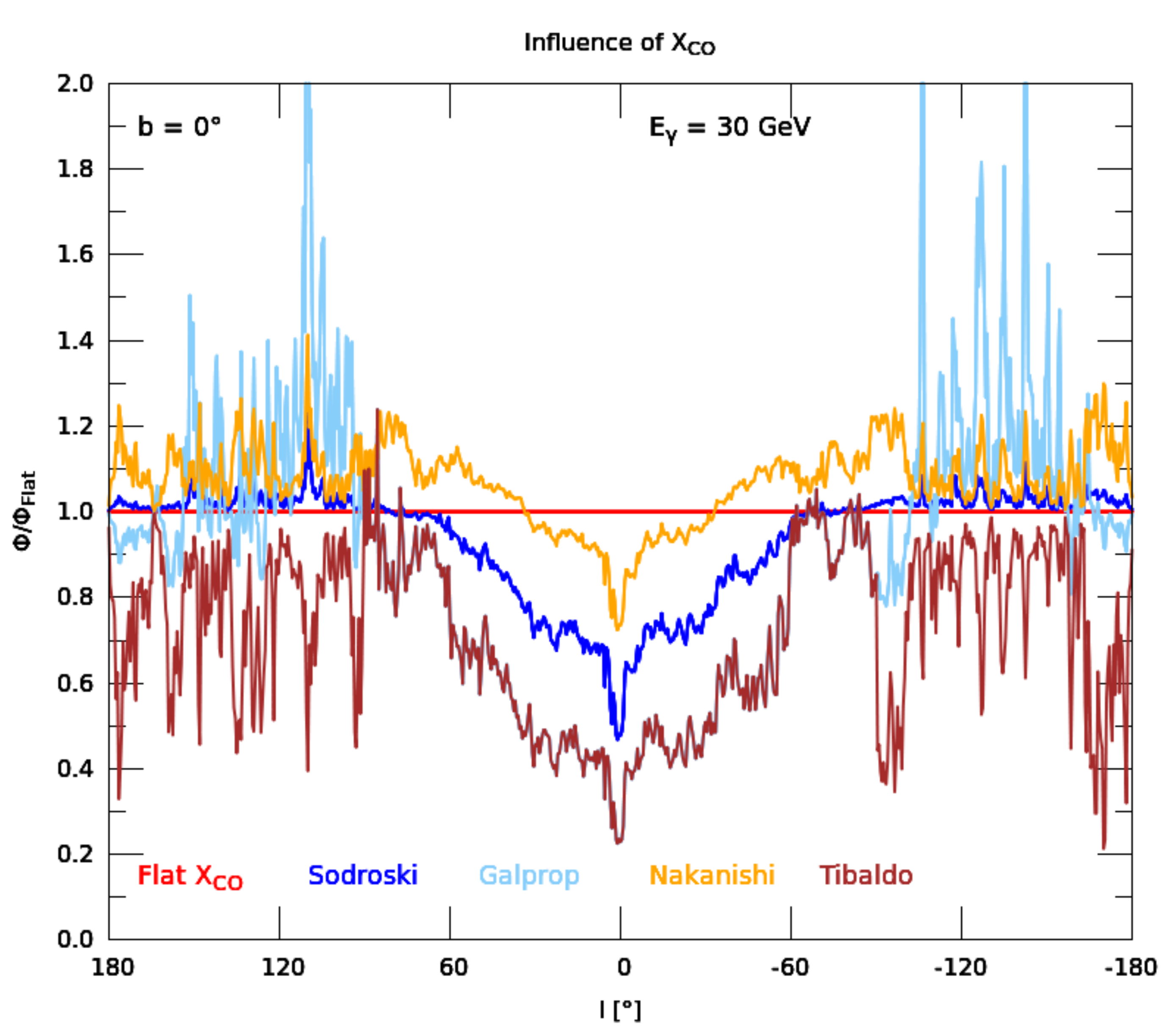}
\caption{
The various X$_{\rm CO}$ Galactic radial profiles available in the
literature are displayed in the left panel.
The corresponding variations in the $\gamma$-ray flux at 30~GeV and
$b = 0^{\circ}$ are displayed relative to our reference model of
Fig.~\ref{fig:ref_map} for which the constant value of
$2.3 \times 10^{20}$ molecules cm$^{-2}$ (K.km.s$^{-1}$)$^{-1}$
has been assumed for X$_{\rm CO}$.
The three (cyan) peaks, whose maxima have not been displayed for clarity,
reach a value of 3.0, 2.3, and 2.3 respectively from left to right.
}
\label{fig:XCO}
%
\vskip -0.25cm %
%
\end{figure}

Most of hydrogen gas in the Galactic plane is under the molecular
form H$_2$. However, directly measuring its column density is extremely
difficult as this molecule is homopolar. It does not shine in the
infrared where its spectroscopy is rather poor. Conditions in molecular
clouds are furthermore unsuitable for the rotational transitions of H$_2$.
Carbon monoxide (CO), the most abundant component of interstellar gas
after H$_2$ and the almost undetectable helium, has a non vanishing
dipole moment with rotational transitions which are easily excited under
molecular cloud conditions.
The question is then to determine the relative abundance of H$_2$ with
respect to CO. The observed CO intensity, expressed as an integrated
brightness temperature, is denoted by $W_{\rm CO}$ and expressed in
units of K km s$^{-1}$. The H$_2$ column density is then expressed as
\beq
N({\rm H_{2}}) \equiv
{\displaystyle \int}_{\rm \!\!\! los} ds \; n_{\rm H_{2}}
\, = \, X_{\rm CO} \times W_{\rm CO} \;\; ,
\eeq
where $X_{\rm CO}$ is the conversion factor.
Our reference model is based on the value of
$2.3 \times 10^{20}$ molecules cm$^{-2}$ (K km s$^{-1}$)$^{-1}$,
see for instance \citet{1988A&A...207....1S} or
\citet{1993ApJ...416..587B}, which we took as constant throughout the Galaxy.
However, as explained by~\citet{2010arXiv1003.1340G} and
\citet{2010arXiv1011.2019S} who uses hydrodynamical simulations,
H$_2$ and CO formation follows different mechanisms. H$_2$ formation
is mainly ruled by the time available to the gas cloud whereas CO
formation highly depends on the carbon and oxygen abundances (metallicity)
and the strength of the background ultraviolet (UV) radiation field that
dissociates CO molecules. These authors have shown that $X_{\rm CO}$ may
vary from
$0.9$ to
$5 \times 10^{20}$ molecules cm$^{-2}$ (K km s$^{-1}$)$^{-1}$
and even over one order of magnitude in extreme molecular cloud conditions.
Because metallicity and UV radiation field are not homogeneous in our Galaxy,
it is even expected that the $X_{\rm CO}$ factor has a spatial dependence.
As a result, we compared our reference model to the various radial profiles of the $X_{\rm CO}$ factor, which can be found in the
literature. In Fig.~\ref{fig:XCO}, following \citet{2004A&A...422L..47S}, 
we used the profiles given by
\citet{1995ApJ...452..262S} and by
\citet{1997ApJ...480..173S}, as well as the profiles that those authors
inferred from
\citet{1997A&A...328..471I} and
\citet{2002A&A...384...33B}. We moreover used the $X_{\rm CO}$
radial profile, which can be found in the GALPROP package.
We finally implemented the recent studies by
\citet{2006PASJ...58..847N} and modified the GALPROP function
according to the studies of \citet{2009arXiv0907.0312T} for the
second quadrant and \citet{2010arXiv1012.0455T} for the third one.
All these profiles are displayed in the left panel of Fig.~\ref{fig:XCO}.

%

\vskip 0.1cm
As featured in the right panel of Fig.~\ref{fig:XCO}, which concentrates
on the Galactic plane, the variations in the $\gamma$-ray flux, induced
by a change in the $X_{\rm CO}$ factor with respect to our reference
value of
$2.3 \times 10^{20}$ molecules cm$^{-2}$ (K km s$^{-1}$)$^{-1}$,
are very important and may reach up to 200\%. This is probably the largest
source of uncertainty in this work.
At higher latitudes, where atomic hydrogen HI starts to dominate
molecular hydrogen H$_2$ -- however the CO
survey by \citet{2001ApJ...547..792D} is at that time only fragmentary as it
covers the limited band in latitude $- 30^{\circ} \leq b \leq 30^{\circ}$
-- the
uncertainties due to the $X_{\rm CO}$ factor drop very fast. Moreover,
one should stress that only the radial variations of $X_{\rm CO}$ have
been considered here, although we expect the UV radiation field to vary with
the distance $|z|$ to the Galactic plane. So should $X_{\rm CO}$, but no
description of this effect has been published so far.

In the recent years some observations have led to the conclusions \citep{2005Sci...307.1292G,2010ApJ...710..133A,2011arXiv1101.2029P} that not only the CO to H$_2$ relation is not completely understood but even worse, both gas distributions are sometimes uncorrelated: indeed there are molecular clouds of \textit{dark gas} that contain copious amounts of molecular hydrogene but no CO at all. The precise amount of \textit{dark gas} in the Galaxy is not known yet, and there is no way to take it into account in this study, hence another source of uncertainties that can be quite large, especially in the direction of the Galactic anti-center.

%
\section{Conclusions}
\label{sec:conclusions}

In this article, we have explored various sources of uncertainty
that affect any determination of the hadronic component of the
Galactic $\gamma$-ray diffuse emission in the GeV-TeV range.
To do so, we used the \citet{Maurin:2001sj} semi-analytic 
two-zone Galactic CR propagation model. The derivation of the densities
of CR protons and helium nuclei everywhere inside the Galactic DH is
immediate once the spectra of these species have been measured at the
Earth. Our method allows therefore a rapid calculation of the $\gamma$-ray
flux and a minute of CPU time is enough to get, on an ordinary PC, a
full sky map at the precision of half a degree both in latitude
and longitude. We devised an efficient tool with which we 
investigated several potential sources of uncertainty and we
quantified the corresponding errors on the $\gamma$-ray flux. These
are more or less important depending on the process at work. Most of
them -- but not all of them -- are likely to be considerably reduced
in the future when better data are available.

\vskip 0.1cm
To commence, the uncertainties on the $\gamma$-ray flux that arise
from a particular choice for the photon production cross sections and
the primary CR proton and helium fluxes are gauged by the spread of
the various curves presented in
Figs.~\ref{fig:local_emissivity} and \ref{fig:emis_p_alpha_fluxes}.
In the case of cross sections, the uncertainty is featured by the band
lying between the long-dashed blue (KNT) and red solid (reference) curves
of Fig.~\ref{fig:local_emissivity}.
It increases from 13\% at 100~MeV up to a maximum of 54\% at 4.5~GeV.
It then decreases down to 20\% at 60~GeV to reach a plateau of 30\% at
1~TeV.
We also showed that the proton and helium fluxes at the Earth's position are key
inputs in the calculation of the $\gamma$-ray flux. A better determination
of the spectra of these species at high energy would be invaluable. Given
the current observations, our reference model prediction could be varied
by $\pm$ 40\% at 1~TeV.
A better understanding of the nuclear processes at stake and refined
measurements of the proton and helium spectra at high-energy are therefore
mandatory in order to calculate more reliably the Galactic $\gamma$-ray
diffuse emission.
There is actually matter for some hope: colliders like the LHC will soon
explore proton-proton and nucleus-nucleus interactions at unprecedented
energies. The photon production cross sections of these processes are
likely to be better parameterized in the future leading to improvement in
our understanding of the mechanisms at work. The CR proton and helium
fluxes at the Earth will also be soon measured with much better
accuracy by experiments like AMS02 \citep{Incagli:2010zz} or CREAM
\citep{Ahn:2010gv}.

\vskip 0.1cm
We then found that the normalization $K_{0}$ and the spectral index
$\delta$ of the space diffusion coefficient $K$ have little influence
on the $\gamma$-ray flux, especially at high energies. This
surprising result can be understood easily. The transport of high-energy
CR protons and $\alpha$ particles inside the Galactic DH is dominated by
space diffusion. In this regime, the densities of these species are
merely proportional to the ratio of the Galactic production rates
$Q_{\rm tot}$ to the diffusion coefficient $K$. Once the CR proton and
helium fluxes at the Earth are fixed, a variation in $K$ translates
into the same variation in $Q_{\rm tot}$. The distribution of CR protons
and helium nuclei throughout the DH remains unaffected. So does the
$\gamma$-ray diffuse emission.
What actually matters is the competition between convection and diffusion.
When convection is taken into account, with a nonvanishing value for
the wind $V_{C}$, CR primaries are washed away from the Galactic plane
as they diffuse from the molecular ring to the Earth. Their densities
are no longer proportional to the ratio ${Q_{\rm tot}}/{K}$. At fixed
proton and helium densities at the Earth, an increase in $V_{C}$ with
respect to $K$ translates into brighter sources and above all
into a denser population of CR primaries around the molecular ring.
The $\gamma$-ray diffuse emission becomes brighter towards the Galactic
center. It also fades in the direction of the anti-center where the cosmic
ray gradient steepens.
Finally, a change in $K$ or $V_{C}$ only has a moderate effect
on $\Phi_{\gamma}$. Even at sub-GeV energies where the competition between
convection and diffusion is significant, the effect does not exceed 10\%.

\vskip 0.1cm
We also explored how the size of the DH could affect the photon flux.
A radial extension of the DH has a zero-to-moderate effect depending on
the population of supernova remnants at large Galactocentric distances.
Even in the case of the P90 profile and its substantial amount of sources
beyond a radius of 20~kpc, the effect does not exceed 6\% as shown in the
right panel of Fig.~\ref{fig:RGAL}. The $\gamma$-ray diffuse emission is
not at all affected if the L04 profile is now chosen.
This is not the case when the thickness of the DH is varied. The latter
acts as a reservoir inside which the hydrogen is illuminated by CR
protons and $\alpha$ particles. An increase in $L$ translates
directly into a larger abundance of gas with which CR primaries can
interact and hence into a larger photon flux. The effect is significant as
illustrated in Fig.~\ref{fig:L_rel} with a variation of $-30$\% to
$+12$\% around the reference value of the MED model.
This situation is likely to be remedied in the future. Better B/C data
should help for improving the determinations of the ${K}/{L}$ ratio and
of the spectral index $\delta$ as discussed in \citet{Castellina:2005ub}.
Accurate measurements of the ${\rm {^{10}Be} / {^{9}Be}}$ ratio would then
help lift the degeneracy between $L$ and $K$. Once again, there is
matter for some hope with forthcoming experiments like AMS02
\citep{Incagli:2010zz} or CREAM \citep{Ahn:2010gv}.

\vskip 0.1cm
This may not be the case for the distribution of primary CR sources along
the Galactic plane. The radial profile of supernova remnants indeed has
a significant effect on the $\gamma$-ray diffuse emission with variations
as large as 50\% towards the Galactic center and 70\% in the opposite
direction. As long as the primary CR source distribution is not
determined better, the $\gamma$-ray flux will be affected by a large
uncertainty. Radio surveys and observations (by the Fermi or HESS
collaborations for instance) of active sources should shed some light
on that problem.
Interestingly enough, our results are insensitive to the different source
profiles at a longitude of about $\pm 60^{\circ}$ as indicated in the left
panel of Fig.~\ref{fig:profil}.

\vskip 0.1cm
We finally showed that the $\gamma$-ray diffuse emission is
extremely sensitive to the Galactic spatial hydrogen distribution.
As expected, predictions depend significantly on the HI and CO maps
in three dimensions, as well as on the assumptions on the $X_{\rm CO}$
conversion factor. Inside the red regions of Fig.~\ref{fig:galprop},
our prediction is twice as large as that of GALPROP.
%
%
The $\gamma$-ray diffuse emission is actually a probe of the gas
distribution inside the Milky Way and offers a unique tool for measuring
the density of hydrogen along the line of sight, hence an extreme
sensitivity to $n_{\rm H}$.
For this tool to work efficiently, all the other sources
of uncertainty need to be reduced. Much effort should be put into a
better determination of the thickness of the DH and the supernova remnant radial
profile.

\vskip 0.1cm
Our conclusions on the effects of CR propagation are qualitatively
comparable to the results of \citet{Cumberbatch:2010ii} although
a quantitative comparison is not possible. These authors have
concentrated on a small region in latitude with
$10^{\circ} \leq |b| \leq 20^{\circ}$ and not on the entire sky.
They vary the half-depth $L$ of the DH from 4 to 11~kpc. This
range is more limited than ours. They have above all explored energies
as low as 20~MeV and incorporated into the $\gamma$-ray flux the
contributions of electron bremsstrahlung and inverse Compton which
are much more uncertain than the hadronic contribution. The spread
in their predictions reaches one order of magnitude at sub-Gev energies.
We plan to investigate that problem in a future publication. We
scrutinized here the various sources of error on the predictions
of the hadronic component that dominates at GeV-TeV energies and
stressed the importance of CR measurements and studies of the astrophysical
sources of the Galactic plane, like supernova remnants.

%

\begin{acknowledgements}
T.D. is really thankful to Luigi Tibaldo for his help on $X_{\rm CO}$ functions.
This work was supported by the Spanish MICINN’s  Consolider-Ingenio 2010
Program under grant CPAN CSD2007-00042. We also acknowledge the support of the
MICINN under grant FPA2009-08958, the Community of Madrid under grant
HEPHACOS S2009/ESP-1473, and the European Union under the Marie Curie-ITN
program PITN-GA-2009-237920.
P.S. expresses his gratitude to the Institut universitaire de France IUF.
This work could not have been done without its help.

\end{acknowledgements}

%
\bibliographystyle{aa}
\bibliography{dfps_gamma_pion}
%

\end{document}